\documentstyle[aps,epsf]{revtex}
\newcommand{\figsize}{10 cm}

\input epsf
\begin{document}

\title{Random matrix model for chiral symmetry breaking and color
  superconductivity in QCD at finite density.\thanks{An animated gif movie
    showing the evolution of the phase diagram with the chiral and diquark
    coupling parameters can be viewed at
    http://www.nbi.dk/\~{}vdheyden/QCDpd.html}}

\author{Beno\^\i t Vanderheyden and A. D. Jackson \\
The Niels Bohr Institute, Blegdamsvej 17, DK-2100 Copenhagen \O, Denmark.}

\date{\today}

\maketitle

\begin{abstract}
  
  We consider a random matrix model which describes the competition between
  chiral symmetry breaking and the formation of quark Cooper pairs in QCD at
  finite density. We study the evolution of the phase structure in
  temperature and chemical potential with variations of the strength of the
  interaction in the quark-quark channel and demonstrate that the phase
  diagram can realize a total of six different topologies.  A vector
  interaction representing single-gluon exchange reproduces a topology
  commonly encountered in previous QCD models, in which a low-density chiral
  broken phase is separated from a high-density diquark phase by a
  first-order line.  The other five topologies either do not possess a
  diquark phase or display a new phase and new critical points.  Since these
  five cases require large variations of the coupling constants away from the
  values expected for a vector interaction, we conclude that the phase
  diagram of finite density QCD has the topology suggested by single-gluon
  exchange and that this topology is robust.

\end{abstract} 

\section{Introduction}

A number of early and recent model studies of finite density
QCD~\cite{Bar77,BaiLov84,AlfRaj98,RapSch98} have suggested that quark Cooper
pairs may form above some critical density and lead to `color
superconducting' matter.  Although perturbation theory performed on
single-gluon exchange suggests pairing gaps of a few MeV~\cite{BaiLov84},
some recent calculations including non-perturbative interactions, either in
the form of the Nambu Jona-Lasinio (NJL)
model~\cite{AlfRaj98,BerRaj99,SchKle99,low}  or those induced by
instantons~\cite{RapSch98,CarDia99}, indicate gaps as large as $100$ MeV.
Theses values imply that color superconductivity may be relevant to the
physics of heavy-ion collisions and neutron stars.

Quark pairing singles out one color direction and thus spontaneously breaks
$SU(3)_{\rm color}$ to $SU(2)_{\rm color}$. The pattern of symmetry breaking
may however be richer, since the formation of condensates in one $\langle qq
\rangle$ channel competes with the breaking of chiral symmetry in the
orthogonal $\langle \bar q q\rangle$ channel.  In an earlier
paper~\cite{VanJac99} we formulated random matrix models for both chiral and
diquark condensations in the limit of two quarks flavors and zero chemical
potential. Our aim was to understand the phase structures which result from
the competition between the two forms of order solely on the basis of the
underlying symmetries. In this spirit, we constructed random matrix
interactions for which the single quark Hamiltonian satisfies two basic
requirements: (1) its block structure reflects the color $SU(N_c)$ and chiral
$SU(2)_{\rm L} \times SU(2)_{\rm R}$ symmetries of QCD and (2) the single
quark Hamiltonian is Hermitian. This last condition ensures the existence of
well-defined relationships between the order parameters and the spectral
properties of the interactions. In particular, condition (2) is obeyed by
single-gluon exchange, which is generally regarded as the relevant
description of QCD. Aside from conditions (1) and (2), the dynamics of the
interactions does not contain any particular structure.  In order to solve
the model exactly, we described this ``dynamics'' by independent Gaussian
distributions of matrix elements.
 
In practice, what distinguishes the interactions from one another is their
respective coupling constants, $A$ and $B$, in the $\langle qq \rangle$ and
$\langle \bar q q \rangle$ channels. The ratio $B/A$ measures the balance
between chiral and diquark condensation forces. The absolute magnitudes of
$A$ and $B$ play a secondary role. They introduce a scale in the condensation
fields but do not affect the phase structure. We have shown
in~\cite{VanJac99} that the condition (2) of Hermiticity forces $B/A$ to be
smaller or equal to $N_c/2$. This constraint results in the absence of a
stable diquark phase in the limit of zero chemical potential.

The purpose of this paper is to extend our previous analysis to non-zero
chemical potentials, for which the interactions cease to be Hermitian.
Non-Hermitian interactions lead to considerable difficulties in numerical
calculations in both lattice and random matrix theories.  Standard Monte
Carlo techniques fail because the fermion determinant in the action is
complex and the sampling weights are no longer positive definite. Obtaining
reliable results thus requires a proper treatment of cancellations between a
large number of terms; a problem which has not yet found a satisfactory
solution~\cite{Biel,Ste96,KogMat83,BarBeh86,KogLom95,BarMor97}.  The
present random matrix models possess exact solutions.  The saddle-point
methods  used in~\cite{VanJac99} to derive the free-energy as a
function of the condensation fields remain valid at finite densities and give
exact results in the thermodynamic limit of matrices of infinite dimensions.
Here, we will use these methods to calculate the thermodynamics quantities as
a function of the condensation fields and deduce the phase diagrams from the
field configurations which maximize the pressure.

We will discover that the pressure function has a simple analytic dependence
which leads to polynomial gap equations. This situation reminds us of the gap
equations obtained in a Landau-Ginzburg theory near criticality, and the
random matrix approach is analogous in many respects.  Both theories
associate a change of symmetry with a change of state in the system.  They
are both mean-field and describe the dynamics of a reduced number of degrees
of freedom $N$, where $N$ scales with the volume of the system. In random
matrix models, these degrees of freedom can be related to the low-lying quark
excitation modes in the gluon background, i.e. the zero modes in an instanton
approximation of that background~\footnote{This relation is explicit in the
  random matrix models where only chiral symmetry is
  considered~\cite{ShuVer93,Ver94,Ver99}; it is expected to remain true once
  color symmetry is also taken into account~\cite{VanJac99}.}. In the
Landau-Ginzburg formulation, the degrees of freedom correspond to the
long-wavelength modes which remain after coarse-graining. There is, however,
an essential difference in the construction of the two theories. A
Landau-Ginzburg theory starts with the specification of an effective
potential for the relevant degrees of freedom. A random matrix model starts
at a more microscopic level with the construction of an interaction which,
once integrated over, produces an effective potential.  The integration can
carry along dynamical constraints and thus restricts the allowed range of
coupling constants.  An example of such restriction is the Hermiticity
condition in~\cite{VanJac99}, which implies that $B/A \le N_c/2$.  This
condition is characteristic of the dynamics of the interactions which we
consider and remains true at finite chemical potential. We will therefore
take it into account in the following.

In this paper, we consider random matrix models in which both chiral and
color condensations can take place and study the resulting phase diagrams in
temperature and chemical potential. We review the model of
Ref.~\cite{VanJac99} and discuss the general form of the pressure as a
function of the condensation fields in Sec.~\ref{s:models}. We then solve the
gap equations and analyze the six topologies that the phase diagram can
assume in Sec.~\ref{s:pd}.  We compare our results with those of QCD
effective models, discuss the possibility of other symmetry breaking patterns
and the extension to the case of non-zero quark current masses in
Sec.~\ref{s:discuss}.  Section~\ref{s:conclusions} presents our conclusions.

\section{Formulation of the model}
\label{s:models}

\subsection{The partition function}
\label{ss:models}

The generalization of the matrix models introduced in Ref.~\cite{VanJac99} to
finite quark densities is straightforward. We represent the quark fields for
each of the two flavors by $\psi_1^{\phantom \dagger}$ and $\psi_2^{\phantom
  \dagger}$. The partition function at temperature $T$ and chemical potential
$\mu$ is then
\begin{eqnarray}
&& Z(\mu,T) = \int {\cal D}H\,{\cal D}\psi_1^\dagger\, {\cal
  D}\psi_1^{\phantom\dagger}
\,{\cal D}\psi_2^T\,{\cal D}\psi_2^* \nonumber \\
&& \times \exp{\left[i \left( \begin{array}{c}
        \psi_1^\dagger \\
        \rule{0pt}{1ex} \\
        \psi_2^T
       \end{array}
\right)^T 
\left( \begin{array}{cc}
          {\cal H} + \left(\pi T+ i \mu\right) \gamma_0 + i m & 
                   {\eta P_\Delta}\\
          \rule{0pt}{1ex} & \rule{0pt}{1ex} \\
          -\eta^* P_\Delta& 
                   -{\cal H}^T + \left(\pi T-i \mu\right) \gamma_0^T - i m 
       \end{array}
\right)
\left( \begin{array}{c} 
       \psi_1^{\phantom\dagger} \\
       \rule{0pt}{1ex} \\
       \psi_2^*
       \end{array}
\right)
\right].} 
\label{partfunc}
\end{eqnarray} 
Here, ${\cal H}$ is a matrix of dimension $4 \times N_c \times N$ which
represents the interaction of a single quark with a gluon background. Its
measure, ${\cal D} H$, will be discussed below.  The parameters $m$ and
$\eta$ select a particular direction for chiral and color symmetry breaking
and are to be taken to zero in the appropriate order at the end of the
calculations.  The current quark mass $m$ is to be associated with the chiral
order parameters $\langle \psi^\dagger_1 \psi_1^{\phantom \dagger} \rangle$
and $\langle \psi^T_2 \psi_2^* \rangle$.  The complex parameter $\eta$ is to
be associated with the order parameter for Cooper pairing, $\langle \psi_2^T
P_\Delta \psi_1^{\phantom \dagger} \rangle$, in which $P_\Delta \equiv i C
\gamma_5 \lambda_2$ ($C$ is the charge conjugation operator) selects the
quark-quark combinations which are antisymmetric in spin and color, i.e. a
chiral isosinglet, Lorentz scalar, and color $\bar 3$ state.  Note that in
order to permit the construction of correlations between the fields $\psi_1$
and $\psi_2^T$, we have transposed the single quark propagator in the second
flavor, hence the upperscript $T$.

The interaction ${\cal H}$ is intended to mimic the effects of gluon
fields and thus explicitly includes the desired chiral and color
symmetries. Of central interest is single-gluon exchange, which has the
chiral block-structure
\begin{eqnarray}
{\cal H_{\rm sge}}& = & \left(
                 \begin{array}{cc}
                    0         & W_{\rm sge} \\
                    W^\dagger_{\rm sge} & 0
                  \end{array}
              \right),
\label{chiralblock}
\end{eqnarray}
where $W_{\rm sge}$ has the spin and color block-structure of a vector
interaction,
\begin{eqnarray}
W_{\rm sge} & = & \sum_{\mu =1}^4 \sum_{a=1}^8 \sigma_\mu^+ \otimes \lambda^a \otimes
A^{\mu a}.
\label{W}
\end{eqnarray}
Here, $\sigma_\mu^+=(1,i\vec  \sigma)_\mu$ are  the $2\times2$ spin matrices,
and $\lambda_a$  denote the $N_c \times  N_c$ Gell-Mann matrices. The $A^{\mu
  a}$ are real $N \times N$ matrices which represent the gluon fields.

Since we want to explore the evolution of the phase structure as the balance
between chiral and diquark condensations is changed, we consider the larger
class of Hermitian interactions to which single-gluon exchange belongs. 
As noted in the introduction, this
choice is motivated by the fact that Hermitian interactions 
have a clear relationship between the order parameters and the
spectral properties, in the form of Banks-Casher formulae~\cite{BanCas80}.  
We write an Hermitian interaction ${\cal H}$ as an expansion into a direct
product of the sixteen Dirac matrices $\Gamma_C$ times the $N_c^2$ color
matrices. The matrix elements are given by
\begin{eqnarray}
{\cal H}_{\lambda i \alpha k;\kappa j \beta l}  & = &
\sum_{C=1}^{16} \left( \Gamma_C^{\phantom T} \right)_{\lambda i;\kappa j} 
\sum_{a=1}^{N_c^2} \Lambda^a_{\alpha\beta} 
\left(A^{C a}_{\lambda\kappa}\right)_{k l},
\label{calH}
\end{eqnarray}
where the indices ($\lambda$,$\kappa$), $(i,j)$, and $(\alpha,\beta)$
respectively denote chiral, spin, and color quantum numbers, while $(k,l)$
are matrix indices running from $1$ to $N$.  The $\Lambda^a$ represent the
color matrices $\lambda^a$ when $a \le N_c^2 -1$ and the diagonal matrix
$(\delta_c)_{\alpha\beta}=\delta_{\alpha\beta}$ when $a = N_c^2$. The
normalization for color matrices is ${\rm Tr}[\lambda^a \lambda^b]=2
\delta_{ab}$ and ${\rm Tr}[\delta_c^2]=N_c$; the normalization of the Dirac
matrices is ${\rm Tr}[\Gamma_C \Gamma_{C'}]=4 \delta_{C C'}$.

The random matrices $A^{Ca}$ are real when $C$ is vector or axial
vector ($C=V,A$) and real symmetric when $C$ is scalar, pseudoscalar, or
tensor $(C=S,P,T)$~\cite{VanJac99}. Their measure is
\begin{eqnarray}
{\cal D} H &=& \left\{\prod_{C a}\prod_{\lambda \kappa} {\cal
D}A^{Ca}_{\lambda\kappa}\right\} 
\exp\left[ - N \sum_{C a}
\sum_{\lambda\kappa}\,\beta_C\Sigma_{Ca}^2 \,\,{\rm Tr}
[A^{Ca}_{\lambda\kappa} (A^{Ca}_{\lambda\kappa})^T] \right],
\label{measure}
\end{eqnarray}
where ${\cal D}A^{Ca}_{\lambda \kappa}$ are Haar measures. Here, $\beta_C=1$
for $C=V,A$ and $\beta_C=1/2$ for $C=S,P,T$. We want to mimic interactions
which in a four-dimensional field theory would respect color $SU(N_c)$ and
Lorentz invariance in the vacuum. Therefore, we choose a single variance
$\Sigma_{Ca}$ for all channels which transform equally under color and space
rotations.

The temperature and chemical potential enter the model in Eq.~(\ref{Omega})
through the inclusion of the first Matsubara frequency in the single quark
propagator. Such $T$ and $\mu$ dependence is certainly oversimplified but none
the less sufficient to produce the desired physics. Our purpose is to
understand the general topology of the phase diagram and not to provide
explicit numbers.  More refined treatments including, for instance, all
Matsubara frequencies would modify the details of the phase diagram and map
every $(\mu,T)$ coordinate to a new one.  However, any such mapping will
necessarily be monotonic and will conserve the topology. We note in
particular that we do not assume a temperature dependence of the variances,
i.e. we neglect the $T$-dependence of the gluon background.  A non-analytic
behavior in $T$ should only arise in the contribution to the thermodynamics
from the degrees of freedom related to chiral and diquark condensations. We
thus expect a realistic $T$-dependence of the gluon background (and of the
variances) to be smooth and not to affect the overall phase topology.

We choose the signs of the $T$- and $\mu$-dependencies to mimic a diquark
condensate which is uniform in time, i.e. which does not contain a proper
pairing frequency. In a microscopic theory formulated in four-momentum space,
the absence of a proper frequency leads to pairing between particle and hole
excitations with energies which are symmetric around the Fermi surface. We
simulate this effect by selecting  opposite $T$-dependences for the fields
$\psi_1^{\phantom T}$ and $\psi^T_2$, while maintaining the same
$\mu$-dependence.

\subsection{The pressure function}
\label{s:pressure}

The integration over the random matrix interactions is Gaussian and can thus
be performed exactly. Following the procedure of Ref.~\cite{VanJac99}, we use
a Hubbard-Stratonovitch transformation to introduce two auxiliary variables
$\sigma$ and $\Delta$, to be associated with the chiral and pairing order
parameters respectively
\begin{eqnarray}
\sigma &\leftrightarrow& 
\langle \psi^\dagger\psi \rangle \equiv 
\langle \psi^\dagger_1 \psi^{\phantom \dagger}_1 \rangle =
- \langle \psi^T_2 \psi^{*}_2 \rangle, \nonumber \\
\Delta &\leftrightarrow& 
\langle \psi \psi \rangle \equiv
\langle \psi^T_2 P_\Delta \psi^{\phantom T}_1 \rangle = 
\left(\langle \psi^\dagger_1 P_\Delta \psi^*_2 \rangle\right)^*.
\label{HS}
\end{eqnarray}
An integration over the fermion fields then reduces the 
partition function to
\begin{eqnarray}
Z(\mu,T) &=& \int d\sigma d\Delta \exp\left[-4 N \Omega(\sigma,\Delta)\right],
\label{partfunceff}
\end{eqnarray}
where $\Omega(\sigma,\Delta)$ is the negative of the pressure,
$P(\sigma,\Delta)$, per degree of freedom and per unit spin and flavor, 
\begin{eqnarray}
\Omega(\sigma,\Delta) & = & - P(\sigma,\Delta) \nonumber \\
& = & A \Delta^2 + B \sigma^2 -{1\over 2} \bigg\{ 
(N_c-2) \log\Big[ \Big( (\sigma + m - \mu)^2 +  T^2 \Big) 
\Big( (\sigma + m + \mu)^2+  T^2 \Big) \Big]  \nonumber \\
&& + 
2 \log\Big[ \Big( (\sigma + m - \mu)^2 +  T^2 + |\Delta+\eta| ^2 \Big) 
\Big( (\sigma + m  + \mu)^2 +  T^2 + |\Delta+\eta|^2 \Big) \Big] 
\bigg\},
\label{Omega}
\end{eqnarray}
where we have dropped the prefactors $\pi$ in the temperature dependence for
simplicity.  Here, the coupling constants $A$ and $B$ are weighted averages
of the Fierz coefficients $f^{Ca}_\Delta$ and $f^{Ca}_\chi$ obtained
respectively by projecting the interaction $\Gamma_C \otimes \Lambda_a$ onto
chiral and diquark channels,
\begin{eqnarray}
A = 2 \left(\sum_{Ca} \Sigma_{Ca}^{-2} f_\Delta^{Ca}\right)^{-1}, \,\,
B = 2 \left(\sum_{Ca} \Sigma_{Ca}^{-2} f_\chi^{Ca}\right)^{-1}.
\label{AB}
\end{eqnarray}
To make contact to microscopic theories, we note that a small coupling
limit in either channel corresponds to a small Fierz constant and hence to
large parameters $A$ or $B$. This limit favors small fields $\Delta$ or
$\sigma$. Because we can always rescale the condensation fields by either
$\sqrt{A}$ or $\sqrt{B}$, the only independent parameter in Eq.~(\ref{Omega})
is the ratio of $B/A$, which by virtue of Eq.~(\ref{AB}) is a measure of the
balance between the condensation forces. Again, Hermitian matrices ${\cal H}$
satisfy $B/A \le N_c/2$. 

The mass $m$ and the parameter $\eta$ explicitly break chiral and color
symmetries. They act as external fields which select a particular direction
for the condensation pattern, and should be taken to zero at the end of the
calculations. (They can also be kept constant to study the effect of a small
external field, a point which we take in Sec.~\ref{s:discuss}.) They are
useful for obtaining the order parameters from derivatives of the partition
function.  In the thermodynamic limit $N \to \infty$, $Z(\mu,T)$ in
Eq.~(\ref{partfunceff}) obeys
\begin{eqnarray}
\lim_{N \to \infty} \log Z(\mu,T) = - \lim_{N \to \infty} 4 N \,\,
{\min}_{\sigma,\Delta} \left\{\Omega(\sigma,\Delta)\right\},
\end{eqnarray}
where the right side represents the global minimum of $\Omega$ in
Eq.~(\ref{Omega}) for fixed $\mu$ and $T$.  The order parameters for chiral and
diquark condensations are given as
\begin{eqnarray}
\langle \psi^\dagger \psi \rangle  =  \lim_{m \to 0} \lim_{N \to \infty} 
\left. \frac{1}{4 N_f N} \frac{\partial \log  Z}{\partial m} 
\right|_{\eta,\eta^*=0}, \quad
\langle \psi^T \psi \rangle  =  - \lim_{\eta,\eta^* \to 0}  \lim_{N\to \infty}
\left. \frac{1}{4 N} \frac{\partial \log Z}{\partial \eta^*}\right|_{m=0},
\label{orderparameters}
\end{eqnarray}
where the number of flavor is $N_f = 2$. Note that the thermodynamic limit $N
\to \infty$ must be taken first before the small field limit $m,\eta \to 0$,
see~\cite{JacVer96}.  Given the $m$- and $\eta$-dependences of the log terms
in Eq.~(\ref{Omega}), we have
\begin{eqnarray}
\langle \psi^\dagger \psi \rangle = B\, \sigma(\mu, T), \quad 
\langle \psi^T \psi \rangle  =  A\, \Delta(\mu,T),
\end{eqnarray}
where $\sigma(\mu,T)$ and $\Delta(\mu,T)$ are the condensation fields
which minimize $\Omega(\sigma,\Delta)$ for fixed $\mu$ and $T$.

Equation~(\ref{Omega}) is the main expression from which we will deduce the
phase structure as a function of the ratio $B/A$.  Its form is very simple to
understand. The quadratic terms correspond to the energy cost for creating
static field configurations with finite $\sigma$ and $\Delta$. The log terms
represent the energy of interaction between the condensation fields and the
quark degrees of freedom.  They can be written in a compact form as ${\rm
  Tr}[\log S(\sigma,\Delta)] = \log {\rm det}~S(\sigma, \Delta)$, where ${\rm
  Tr}$ is a trace in flavor, spin, and color, and $S$ is the single quark
propagator in a background of $\sigma$ and $\Delta$ fields. Substituting the
Mastubara frequency $T \to i p_4$, it becomes clear that the poles of $S$ in
$p_4$ correspond to the excitation energies of the system.  From
Eq.~(\ref{Omega}), we see that two colors develop gapped excitations with
\begin{eqnarray}
p_4 & = & \pm \sqrt{(\sigma \mp \mu)^2+\Delta^2},
\label{gapped}
\end{eqnarray}
where the plus and minus signs respectively correspond to particle and
antiparticle modes. The $N_c-2$ remaining colors have ungapped excitation
with
\begin{eqnarray}
p_4 & = & \pm |\sigma \mp \mu|.
\label{ungapped}
\end{eqnarray}

It is important to recognize that the potential $\Omega$ in Eq.~(\ref{Omega})
contains the contribution to the thermodynamics from only the low energy
modes of Eqs.~(\ref{gapped}) and (\ref{ungapped}).  The right side of
Eq.~(\ref{Omega}) thus corresponds to the non-analytic piece in the
thermodynamic potential which describes the critical physics related to
chiral and color symmetry breaking. In a microscopic model, the right side of
Eq.~(8) would also contain a smooth analytic component $\Omega_{\rm
  reg}(\mu,T)$ which arises from all other, non-critical degrees of freedom
in the system. Our model does not contain this contribution and thus cannot
be taken as a quantitative description of bulk thermodynamics properties; the
model is constructed to describe the critical properties and should be used
as such.

We show in the next section that the forms of the potential
$\Omega(\sigma,\Delta)$ in Eq.~(\ref{Omega}) and of the associated excitation
energies in Eqs.~(\ref{gapped}) and (\ref{ungapped}) are sufficient to
produce a rich variety of phase diagrams which illustrate in a clear way the
interplay between chiral and color symmetries. In particular, we will find
that single-gluon exchange reproduces the topology obtained in many
microscopic models of finite density QCD, see for
instance~\cite{RapSch98,CarDia99,BerRaj99,SchKle99}.

\section{Exploring the phase diagrams}
\label{s:pd}

We noted earlier that the potential $\Omega$ in Eq.~(\ref{Omega}) is very
similar to a Landau-Ginzburg functional.  Although $\Omega(\sigma,\Delta)$ is
not an algebraic function, it is equivalent to a polynomial of order
$\sigma^6$ and $\Delta^4$: The gap equations reduce to coupled polynomial
equations of fifth order in $\sigma$ and third order in $\Delta$.
Generically, four types of solutions exist:
\begin{itemize}
\item (i) the $0$-phase, the trivial phase in which both
$\sigma$ and $\Delta$ vanish, 
\item (ii) the $\chi$-phase, in which chiral symmetry
is spontaneously broken but $\Delta=0$,
\item  (iii) the $\Delta$-phase, in which
color symmetry is spontaneously broken but $\sigma=0$, 
\item (iv) the $\chi\Delta$
phase, a mixed broken symmetry phase in which both fields are
non-vanishing, $\sigma \neq 0$ and $\Delta \neq 0$.  
\end{itemize}
The $\chi\Delta$-phase is thermodynamically distinct from the $\chi$- and
$\Delta$-phases and is not a ``mixture'' of these two phases.

\begin{figure}[htb]
\epsfxsize=\figsize
\centerline{
\epsffile{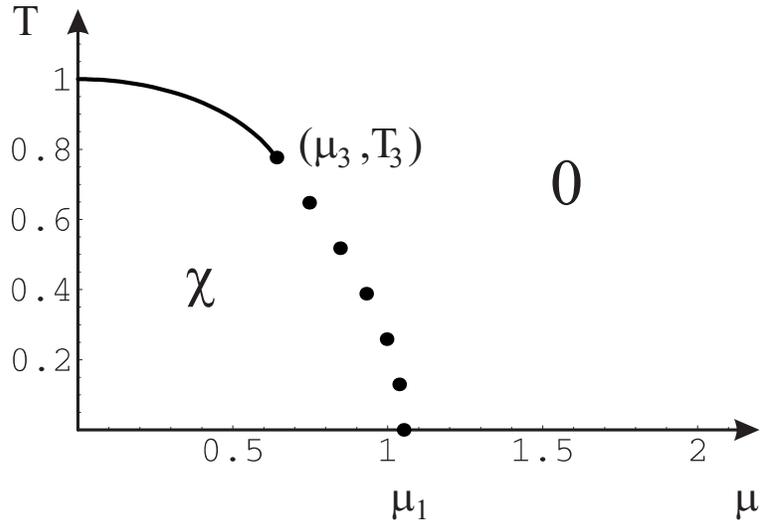}}
\vspace{.5cm}
\caption{
  Phase diagram for $B/A \le 0.418$ and $N_c=3$, as a function of the quark
  chemical potential $\mu$ and the temperature $T$. For this and all the
  following figures, continuous curves represent second-order lines while
  first-order lines are plotted with dots. Here, $\chi$ is the chiral phase
  and $\Delta$ is the diquark phase. First- and second-order lines join
  tangentially at the tricritical point $(\mu_3,T_3)$.}
\label{f:panel1}
\end{figure}

\begin{figure}[hb]
\epsfxsize=\figsize
\centerline{
\epsffile{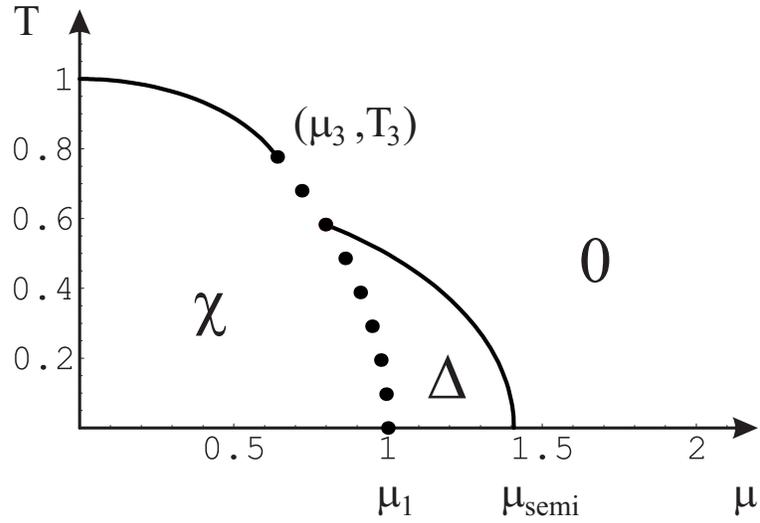}}
\vspace{0.5cm}
\caption{Phase diagram for the coupling ratio realized by single-gluon 
  exchange, $B/A=0.75$. The transition from the chiral phase $\chi$ to the
  diquark phase $\Delta$ is first-order while that from the $\Delta$ to the
  trivial phase is second-order.}
\label{f:panel2}
\end{figure}

\begin{figure}[ht]
\epsfxsize=\figsize
\centerline{
\epsffile{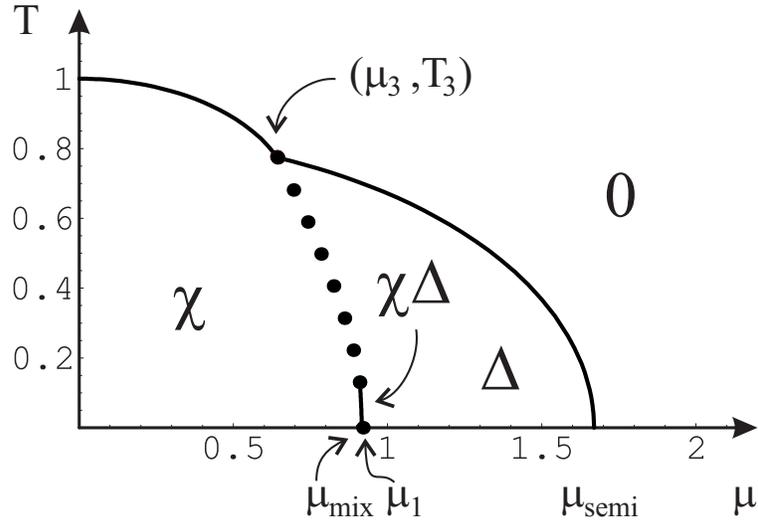}}
\vspace{.5cm}
\caption{ Phase diagram for a ratio $B/A=1.054$. The mixed broken symmetry
phase $\chi\Delta$ emerges out of the chiral phase via a second-order
transition at $\mu_{\rm mix}$, and undergoes a first-order transition
towards the diquark phase at $\mu_{1}$.}
\label{f:panel3}
\end{figure}

\begin{figure}[hb]
\epsfxsize=\figsize
\centerline{
\epsffile{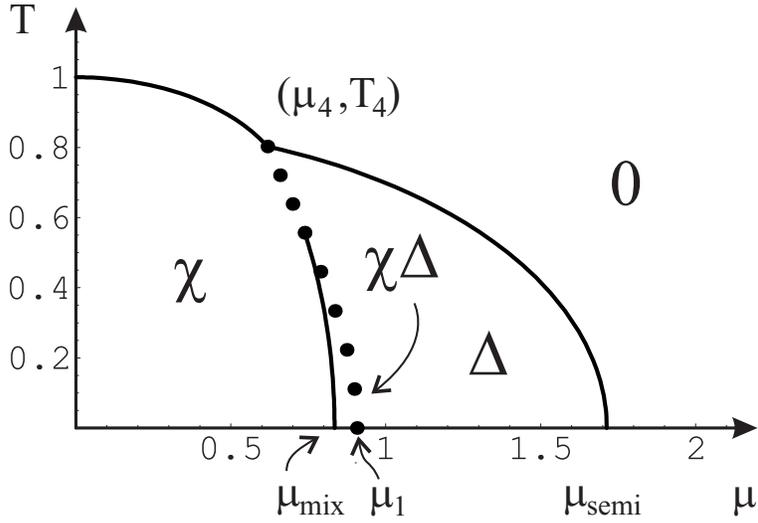}}
\vspace{.5cm}
\caption{
Phase diagram for a ratio $B/A=1.1$. Compared to
Fig.~3,
the tricritical point $(\mu_3,T_3)$ no longer exists.
The trivial, chiral, and diquark phases now meet at the new critical point
$(\mu_4,T_4)$. We note in particular that the first and second-order lines 
bordering the chiral phase do not meet tangentially at $(\mu_4,T_4)$.}
\label{f:panel4}
\end{figure}

\begin{figure}[ht]
\epsfxsize=\figsize
\centerline{
\epsffile{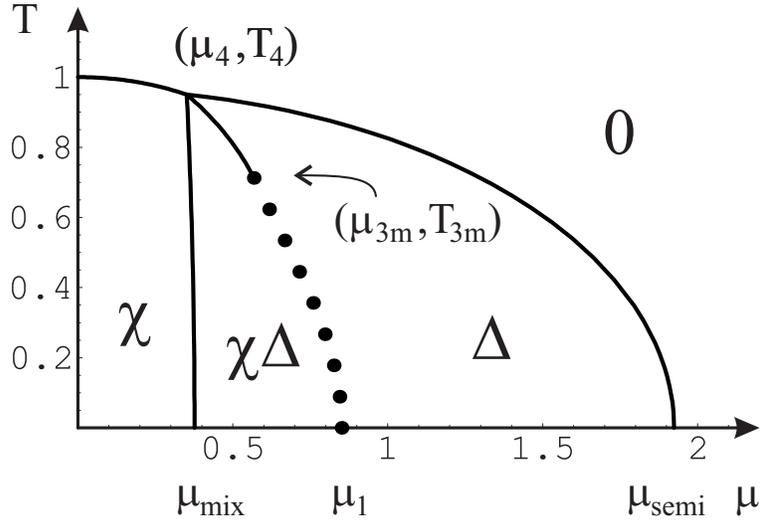}}
\vspace{.5cm}
\caption{Phase diagram for a ratio $B/A=1.4$. Compared to
  Fig.~4, $(\mu_4,T_4)$ has become a tetracritical point, at the
  intersection of the four phases. There is also a new tricritical point
  $(\mu_{3\rm m},T_{3\rm m})$ where the first- and second-order line
  separating the $\chi\Delta$- and the $\Delta$-phases join tangentially.}
\label{f:panel5}
\end{figure}

\begin{figure}[hb]
\epsfxsize=\figsize
\centerline{
\epsffile{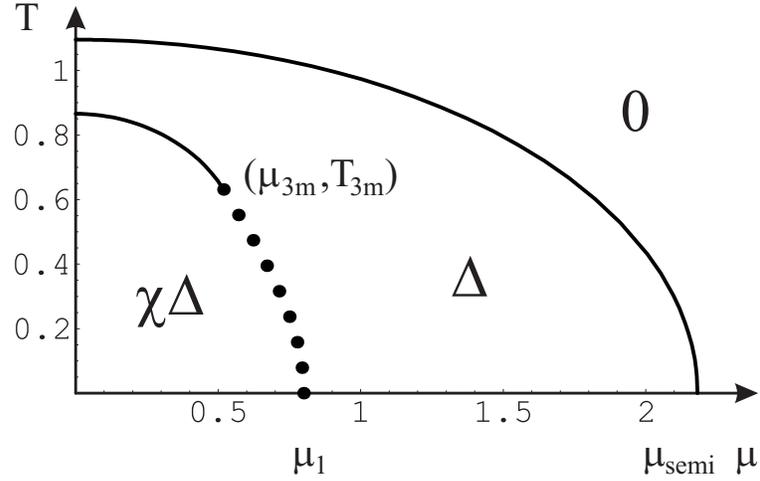}}
\vspace{.5cm}
\caption{Phase diagram for a ratio $B/A=1.8$. This topology is characteristic
  of large ratios $B/A$, which favor diquark over chiral condensation. The
  $\Delta$-phase occupies a large part of the phase diagram, at the expense
  of the mixed broken symmetry phase. The chiral phase has completely
  vanished. We note that since Hermitian interactions realize ratios $B/A <
  1.5$, this last topology actually describes a case out of reach for our
  model.}
\label{f:panel6}
\end{figure}

At a given $\mu$, $T$, and $B/A$, each of the solutions (i) - (iv) can either
be a minimum or a saddle-point of $\Omega$. The complex flow of these
solutions with the variation of $B/A$ leads to large variety of phase
diagrams. The phase structures can however be grouped according to their
topologies as shown in Figs.~\ref{f:panel1}-~\ref{f:panel6}.~\footnote{We
  have rescaled in each figure the units of $T$ and $\mu$  for clarity.}
Figure~\ref{f:panel1} shows the case of smallest values $B/A$, which favor
chiral over diquark condensation. $B/A$ then increases continuously from
Fig.~\ref{f:panel2} to~\ref{f:panel6}. We now discuss Figs.~1-6 in the
six following subsections. We indicate in parentheses the corresponding
ranges of $B/A$ and discuss their limits in the text.

{\bf Pure chiral condensation:} ($0 \le B/A \le 0.139 N_c$ or $0 \le B/A \le
0.418$ for $N_c=3$; see Fig.~1). Generically, the global minimum of $\Omega$
in Eq.~(\ref{Omega}) is realized by field configurations which maximize the
log terms while keeping reasonably low values of the quadratic terms $A
\Delta^2 + B \sigma^2$. We consider first the limit $A \gg B$.  The minimum
of $\Omega$ must then correspond to $\Delta = 0$ in order to avoid the large
energy penalty $\sim A \Delta^2$. No diquark condensation occurs in this
case, and color enters only as a prefactor $N_c$ in the number of degrees of
freedom. This becomes clear if we absorb $N_c$ into $B$ and set $B$ to $1$;
we then recover the potential $\Omega$ studied in the chiral random matrix
models~\cite{HalJac98} which neglect color altogether.

We briefly recall the phase structure in this case.  The gap equation for the
chiral field, $\partial \Omega(\sigma,0)/\partial \sigma = 0$, has a trivial
root $\sigma =0$ and four other roots which satisfy the following quadratic
equation for $\sigma^2$
\begin{eqnarray}
 N_c (\mu^2-T^2)+B (\mu^2+T^2)^2 +
 (2 B (T^2-\mu^2)-N_c) \sigma^2 & + & B \sigma^4 =  0.
\label{sigma4}
\end{eqnarray}
In the high temperature limit $T \gg \mu$, both roots in $\sigma^2$ are
negative. The only real solution of the gap equation is $\sigma=0$, and the
system is in the symmetric phase.  Decreasing $T$ for fixed $\mu$, one
encounters a line of second-order phase transitions
\begin{eqnarray}
L_{\chi,2} \equiv N_c (\mu^2-T^2)+B(\mu^2+T^2)^2 & = & 0,
\label{sigma_second}
\end{eqnarray}
where one of the roots $\sigma^2$ in Eq.~(\ref{sigma4}) vanishes.  Below
$L_{\chi,2}$, the trivial root becomes a local maximum of $\Omega(\sigma,0)$,
and we have a pair of local minima at the real roots $\sigma = \pm \sigma_0$,
where
\begin{eqnarray}
\sigma_0 = \left(\frac{N_c}{2 B}- T^2+\mu^2 + 
\frac{\sqrt{N_c^2-16 B^2 \mu^2 T^2}}{2 B}\right)^{1/2}.
\label{sigma0}
\end{eqnarray}
These roots correspond to a chiral broken phase.  They become
degenerate with $\sigma = 0$ on the second-order line $L_{\chi,2}$, where the
potential $\Omega$ scales as $\Omega(\sigma,0) - \Omega(0,0) \approx
\sigma^4$.  Thus, the critical exponents near $L_{\chi,2}$ are those of a
mean-field $\phi^4$ theory.

The second-order line ends at a tricritical point $(\mu_3,T_3)$ at which all
five roots of the gap equation vanish and where
$\Omega(\sigma,0)-\Omega(0,0)\approx \sigma^6$, giving now critical exponents
of a mean-field $\phi^6$ theory. From Eq.~(\ref{sigma4}), this happens when
$2 B (T_3^2-\mu_3^2)-N_c=0$, which with the use of Eq.~(\ref{sigma_second})
gives
\begin{eqnarray}
\mu_3  =  \sqrt{\frac{N_c}{4 B}} \sqrt{\sqrt{2}-1}, \quad \quad
T_3  =  \sqrt{\frac{N_c}{4 B}} \sqrt{\sqrt{2}+1}.
\label{tricritical}
\end{eqnarray}
For $\mu > \mu_3$, the transition between the chiral and trivial phases
is first-order and takes place along the line of equal pressure 
\begin{eqnarray}
L_{\chi,1} \equiv && \frac{N_c}{2} \left(1+\sqrt{1- 16 \mu^2 T^2
    \frac{B^2}{N_c^2} } -\log \left[\frac{N_c^2}{2 B^2} \left( 1+ \sqrt{1-16
        \mu^2 T^2 \frac{B^2}{N_c^2}} \right) \right] \right)
\nonumber \\
&& + N_c \log\left[\mu^2+T^2\right] + B (\mu^2-T^2) = 0.
\label{Lchi1}
\end{eqnarray} 
This line intercepts the $T=0$ axis at 
$\mu=\mu_1$ which obeys
\begin{eqnarray}
1+{B \mu_1^2\over N_c}+\log({B \mu_1^2 \over N_c})=0.
\label{mu1}
\end{eqnarray}
This gives $\mu_1 = 0.528 \sqrt{N_c/B}$, or $\mu_1 = 0.914/\sqrt{B}$ for
$N_c=3$.

$L_{\rm \chi,1}$ in Eq.~(\ref{Lchi1}) is a triple line.  To see this and
clarify the character of $(\mu_3,T_3)$, it is useful to consider the effect
of a non-zero quark current mass $m$~\cite{HalJac98}.  A mass $m \neq 0$
selects a particular direction for chiral condensation. If we now consider
the three-dimensional parameter space $(\mu,T,m)$, the region delimited by
$L_{\chi,2}$ and $L_{\chi,1}$ in the plane $m=0$ thus appears to be a surface
of coexistence of the two ordered phases with the chiral fields $\pm
\sigma_0$ of Eq.~(\ref{sigma0}).  Along $L_{\chi,1}$, this surface meets two
other `wing' surfaces which extend symmetrically into the regions $m >0$ and
$m <0$.  Each of the wings is a coexistence surface between one of the
ordered phases whose chiral field continues to $\sigma = \pm \sigma_0$ as
$m\to 0$ and the high temperature phase.  Hence, $L_{\chi,1}$ marks the
coexistence of three phases and is a triple line.  The three phases become
identical at $(\mu_3,T_3)$, which is thus a tricritical point. The second-
and first-order lines $L_{\chi,1}$ and $L_{\chi,2}$ join tangentially at
$(\mu_3,T_3)$, see~\cite{DomGre84}.

The onset of diquark condensation modifies the topology that we have just
described. This takes place for coupling ratios $B/A \ge 0.139 N_c$ to which
we now turn.

{\bf The QCD case:} ($0.139 N_c \le B/A \le \alpha_1(N_c)$ or $0.418 \le B/A
\le 1.05$ for $N_c=3$; see Fig.~2). To understand the conditions for the
onset of diquark condensation, consider a pure diquark phase by setting
$\sigma=0$ in Eq.~(\ref{Omega}). The gap equation $\partial \Omega/\partial
\Delta=0$ then has three solutions: $\Delta = 0$, and $\Delta=\pm\Delta_0$
where $\Delta_0 = \sqrt{2/A -\mu^2-T^2}$. In the high $T$ and $\mu$ phase,
only the trivial root is real and the system is in the symmetric phase.
Inside the semicircle
\begin{eqnarray}
L_{\rm \Delta,2} \equiv 2/A-\mu^2-T^2=0,
\end{eqnarray} 
the trivial root becomes a maximum of $\Omega(0,\Delta)$, and we have a pair
of two real minima $\Delta = \pm \Delta_0$ for which color symmetry is
spontaneously broken.  The roots $\pm \Delta_0$ go continuously to zero as
one approaches $L_{\Delta,2}$, which is thus a second-order line provided no
other phase develops. The fate of other ordering forms depends on the
pressure in the diquark phase,
\begin{eqnarray}
P_\Delta=-\Omega_\Delta & = & A (\mu^2+T^2) - 2  + (N_c-2) \log\left( \mu^2 + 
T^2 \right)+ 2 \log\left(\frac{2}{A}\right).
\end{eqnarray}
The maximum pressure $P_\Delta$ is reached on $L_{\Delta,2}$, where it
also equals the pressure of the trivial phase.  The condition for the
onset of diquark condensation is then clear: The semicircle must lie
in part outside the region occupied by the chiral phase. Then, the
maximum pressure in the diquark phase is higher than that of the
chiral phase, and the diquark phase is stable near the semicircle.

To find the minimum ratio $B/A$ for which this condition holds, we compare
the dimensions of the chiral phase to the radius of the semicircle, $\mu_{\rm
  semi}\equiv \sqrt{2/A}$.  When $B/A \ll 1$, the line $L_{\Delta,2}$ lies
well inside the boundaries of the chiral phase, whose linear dimensions are
$T=\sqrt{N_c/B}$ (along $\mu=0$) and $\mu_1=0.528 \sqrt{N_c/B}$ (along
$T=0$). The onset of diquark condensation requires therefore $B/A$ to be
large enough that the semicircle crosses the first-order line between chiral
and trivial phases. This takes place along $T=0$ when $\mu_{\rm
  semi}=\mu_{1}$, or
\begin{eqnarray}
\sqrt{2/A} = 0.528 \sqrt{N_c/B},
\end{eqnarray}
which gives $B/A = 0.139 N_c$, or $B/A = 0.418$ for $N_c =3$.

\begin{figure}[hb]
\epsfxsize=\figsize
\centerline{
\epsffile{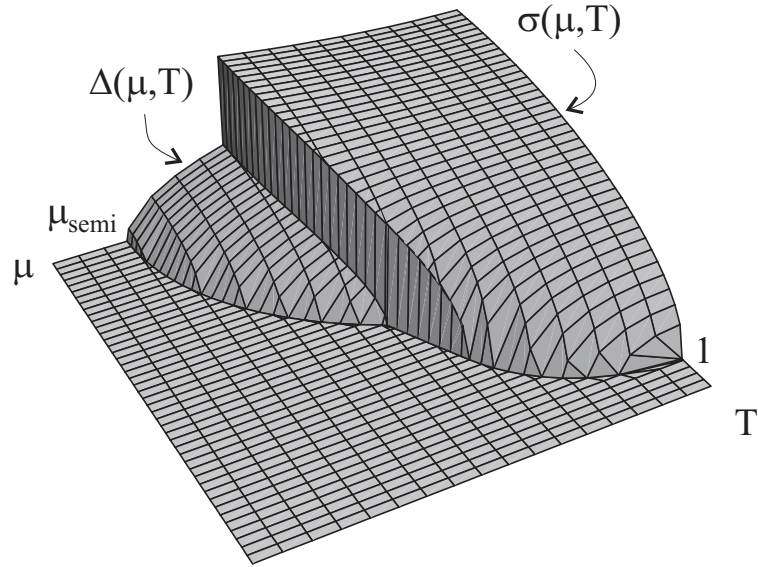}}
\caption{Order parameters for the coupling ratio of single-gluon exchange,
  $B/A=0.75$. The corresponding phase diagram is shown in Fig.~2.  Chiral and
  diquark fields vanish continuously along the second-order lines.}
\label{f:fields}
\end{figure}

For larger values of $B/A$, the diquark phase exists in a region delimited by
$L_{\Delta,2}$, on which it coexists with the symmetric phase, and by a
first-order line of equal pressure with the chiral phase.  This phase diagram
is realized for ratios including that corresponding to single-gluon exchange,
$B/A = N_c/ (2 (N_c-1))$ (or $3/4$ for $N_c=3$), which is the ratio taken in
Fig.~\ref{f:panel2}. We have verified that the two segments of first-order
lines, between the chiral and the diquark phases on the one hand and the
chiral and trivial phases on the other hand, join tangentially.
Figure~\ref{f:fields} shows the chiral and diquark fields as a function of
$\mu$ and $T$. It is worth noting that the chiral field vanishes with a
square root law near the second-order line $L_{\rm \chi,2}$ and is
discontinuous along the first-order line $L_{\chi,1}$.  The diquark field
vanishes with a square root law all along the semicircle line.

The topology in Fig.~\ref{f:panel2} can also be summarized by a simple
counting argument.  Consider the difference between the pressures in the
diquark and the chiral phases along the $T=0$ axis,
\begin{eqnarray}
\Delta \Omega = \Omega_\chi-\Omega_\Delta 
= N_c-2 +(A+B)\mu^2- 2\log \frac{A \mu^2}{2} + N_c \log\frac{B \mu^2}{N_c}.
\label{DeltaOmega}
\end{eqnarray}
This difference varies from $\Delta \Omega = N_c (1 + (2B)/(N_c A)
+\log[(2B)/(N_c A)])$ as $\mu \to \sqrt{2/A}$, to $\Delta \Omega = (N_c-2)
\log \mu^2$ as $\mu \to 0$. We have just argued that $\Delta\Omega$ can be
negative in the former limit if $B/A> 0.139 N_c$. In the latter limit,
however, $\Delta\Omega$ is always negative for $N_c > 2$ and the system is
necessarily in the chiral phase. That the diquark phase must be metastable
for small $\mu$ is clearly a consequence of the fact that chiral condensation
uses all $N_c$ colors while diquark condensation uses only colors $1$ and
$2$. This counting argument was mentioned in early models of color
superconductivity~\cite{AlfRaj98} and is here completely manifest.  For $N_c
> 2$, the diquark phase must appear at densities higher than those
appropriate for the chiral phase.

{\bf Onset of the mixed broken symmetry phase:} ($\alpha_1(N_c) \le B/A \le
N_c/\sqrt{8}$ or $1.05 \le B/A \le 1.06$ for $N_c=3$; see Fig.~3). The chiral
solution $\sigma_0$, Eq.~(\ref{sigma0}), ceases to be a local minimum of
$\Omega(\sigma,0)$ for sufficiently large $\mu$.  It turns into a saddle
point along a line where the second derivative $\partial^2
\Omega/\partial \Delta^2$ vanishes,
\begin{eqnarray}
L_{\rm mix} \equiv 
\left(N_c A - 2 B\right) \left( N_c+\sqrt{N_c^2-16 B^2 \mu^2 T^2}\right)
- 8 B^2 \mu^2 = 0.
\label{mixline}
\end{eqnarray}
A new minimum of $\Omega(\sigma,\Delta)$ with both $\sigma \neq 0$ and
$\Delta \neq 0$ develops in the region to the right of $L_{\rm mix}$.
This new minimum corresponds to a new phase, the $\chi\Delta$-phase, which
competes with the diquark phase. When the $\chi\Delta$-phase first appears
along $L_{\rm mix}$, it has the same pressure as the chiral phase.
Therefore, the condition for this new phase to realize the largest pressure
is that the instability line $L_{\rm mix}$ lies in the region spanned by the
chiral phase. Then, along $L_{\rm mix}$, the new phase necessarily has a
pressure that exceeds that in the diquark phase and is favored.

\begin{figure}[ht]
\epsfxsize=\figsize
\centerline{
\epsffile{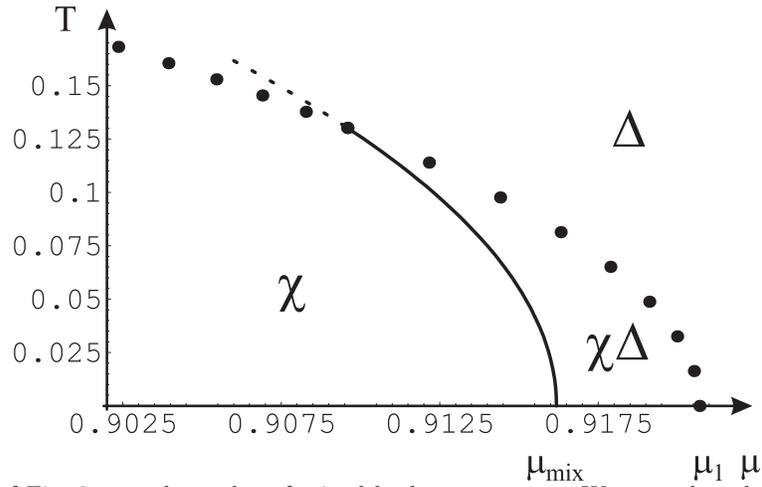}
}
\caption{Expanded view of Fig.~3 near the wedge of mixed broken symmetry.
  We note that the second-order line between the $\chi$- to the $\chi\Delta$
  phase does not meet tangentially with the first-order line between the
  $\chi\Delta$- and the $\Delta$-phases, as indicated by the short dashed
  line.}
\label{f:panel3b}
\end{figure}

\begin{figure}[hb]
\epsfxsize=\figsize
\centerline{
\epsffile{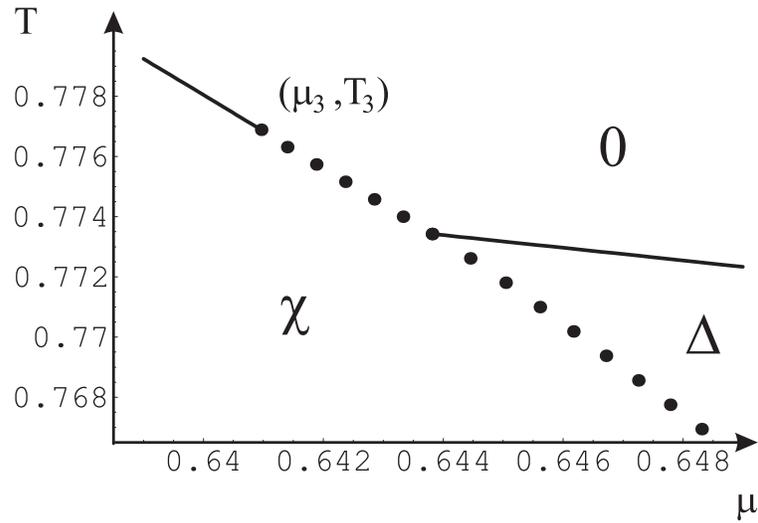}
}
\caption{
  Expanded view of Fig.~3 near the tricritical point $(\mu_3,T_3)$, showing
  that there is a very short first-order segment between the chiral and
  trivial phases. }
\label{f:panel3c}
\end{figure}

To see what ratios $B/A$ are needed for the mixed broken symmetry phase,
consider the point where $L_{\rm mix}$ meets the $T=0$ axis,
\begin{eqnarray}
\mu_{\rm mix}&=& \sqrt{\frac{N_c}{2 B} \left({N_c A\over 2 B}-1\right)}.
\label{mumix}
\end{eqnarray}
When $B/A \ll 1$, $\mu_{\rm mix} \sim \sqrt{N_c/B}/ \sqrt{ 4B/A} \gg
\sqrt{N_c/B}$ and $L_{\rm mix}$ lies well outside the chiral phase.  $L_{\rm
  mix}$ then moves towards the $\chi$-phase as $B/A$ increases. It first
crosses the first-order line between chiral and diquark phases when $\mu_{\rm
  mix}=\mu_1$, where $\mu_1$ is the point of equal pressure which obeys
$\Delta\Omega=0$ in Eq.~(\ref{DeltaOmega}). This condition gives a ratio $B/A
= \alpha_1(N_c)$, where $\alpha_1$ is a non-trivial function of $N_c$ which
we study in the appendix.  Here, we note only that $\alpha_1(3)=1.05$ for
three colors.  When $B/A > \alpha_1(N_c)$, we find that the mixed broken
symmetry phase develops in the wedge bordered by the second-order line
$L_{\rm mix}$ of Eq.~(\ref{mixline}), and a first-order line on which it
coexists with the diquark phase. The phase diagram is shown in
Fig.~\ref{f:panel3}, while an expanded view of the wedge of mixed broken
symmetry and of the first-order line near $(\mu_3,T_3)$ are
respectively shown in Figs.~\ref{f:panel3b} and~\ref{f:panel3c}.

{\bf A new critical point:} ($N_c/\sqrt{8} \le B/A \le \alpha_2(N_c)$ or
$1.06 \le B/A \le 1.163$ for $N_c = 3$; see Fig.~4).  As $B/A$ increases above
$\alpha_1(N_c)$, the semicircle $2/A = \mu^2+T^2$ grows relative to the
chiral phase, and eventually reaches the tricritical point $(\mu_3,T_3)$ of
Eq.~(\ref{tricritical}) when $B/A = N_c / \sqrt{8}$.  At this stage, the
segment of first-order line between chiral and trivial phases disappears. For
$B/A > N_c/\sqrt{8}$, the semicircle meets the second-order line between
chiral and trivial phases, $L_{\chi,2}$ in Eq.~(\ref{sigma_second}), at a new
critical point
\begin{eqnarray}
\mu_4 = \sqrt{{1\over A}\left(1 - {2 B \over N_c A}\right)}, \quad
T_4 = \sqrt{{1\over A}\left(1 + {2 B \over N_c A}\right)},
\label{tetracritical}
\end{eqnarray}
which now separates three phases as shown in Fig. \ref{f:panel4}.

{\bf Coexistence of four phases:} ($\alpha_2(N_c) \le B/A \le N_c/2$ or
$1.163 \le B/A \le 1.5$ for $N_c=3$; see Fig.~5). With still higher coupling
ratios $B/A$, the wedge of mixed broken symmetry in Fig.~\ref{f:panel4} grows
in size relative to the other phases.  Its tip reaches the new critical point
$(\mu_4,T_4)$ when $B/A$ satisfies the condition
\begin{eqnarray}
{B \over A}  = \alpha_2(N_c) \equiv \frac{4 N_c - N_c^{3/2} 
\sqrt{2 N_c-4}}{4 (4 - N_c)},
\label{alpha2}
\end{eqnarray}
the derivation of which we detail in the appendix. We just note here that
$\alpha_2( 3 ) = 1.163$.  This coupling ratio also marks the appearance of
two new critical points, as illustrated in Fig.~\ref{f:panel5}. First, when
$B/A > \alpha_2(N_c)$, the point $(\mu_4,T_4)$ characterizes the coexistence
of all four phases, and it has thus become a {tetracritical} point. The
pressure of the system at $(\mu_4,T_4)$ has the form
$\Omega(\sigma,\Delta)-\Omega(0,0) \approx a \sigma^4 + b \Delta^4+ c
\sigma^2 \Delta^2$, where $a$, $b$, and $c$ are constants detailed in the
appendix.  We note in particular that none of the four second-order lines in
Fig.~5 join tangentially at the tetracritical point.

The second point appearing above $B/A=\alpha_2(N_c)$ is a tricritical
point $(\mu_{\rm 3m},T_{\rm 3m})$ which lies on the boundary between the
$\chi\Delta$- and the $\Delta$-phases. Its origin can be understood from the
similarity between the characters of the mixed broken symmetry phase and the
chiral phase. The diquark field in the $\chi\Delta$-phase is 
\begin{eqnarray}
\Delta(\sigma) & =& \left({1+\sqrt{1+4 A^2 \mu^2 \sigma^2}\over A}
-\mu^2-T^2-\sigma^2 \right)^{1/2}.
\end{eqnarray}
Substituting $\Delta$ in $\Omega(\sigma,\Delta)$ for this expression, we find
that the gap equation for the chiral field, $\partial
\Omega(\sigma,\Delta(\sigma))/\partial \sigma = 0$, has five roots.  Despite
the fact that $\Delta(\sigma) \neq 0$, the dynamics of these roots as a
function of $\mu$ and $T$ is identical to that of the roots in the pure
chiral condensation case discussed at the beginning of this section.  In
particular, the transition to the $\Delta$-phase starts out second-order near
$(\mu_4,T_4)$ and takes place along a line where three of the five roots
vanish. This second-order line is thus determined by the condition
\begin{eqnarray}
\left.\frac{d^2 \Omega}{d \sigma^2}(\sigma,\Delta(\sigma))\right|_{\sigma \to
  0}& = & 0,
\label{partial2Om}
\end{eqnarray}
which is the analog of Eq.~(\ref{sigma_second}) and gives 
\begin{eqnarray}
L_{\chi\Delta \to \Delta,2} \equiv 
(N_c-2) (\mu^2-T^2) + \left(B- A(1-A \mu^2)\right)\left(\mu^2+T^2\right)^2=0.
\label{mixtocol}
\end{eqnarray}
In analogy with the pure chiral case, the transition becomes first order at
high $\mu$; second- and first-order segments join tangentially at a 
  tricritical point $(\mu_{3\rm m},T_{3\rm m})$.

We determine the location of this point as follows.  We observe that, as in
 the pure chiral broken case, the potential along the second-order line
$L_{\chi\Delta \to \Delta,2}$ has the form $\Omega(\sigma,\Delta(\sigma)) -
\Omega(0,0) \approx \sigma^4$. (See Eq.~(\ref{partial2Om}).) 
This scaling form  becomes
$\Omega(\sigma,\Delta(\sigma)) -\Omega(0,0) \approx \sigma^6$ at $(\mu_{3\rm
  m},T_{3\rm m})$ where all five roots vanish.
Thus, the tricritical point is located on the line $L_{\chi\Delta \to
  \Delta,2}$ at the point where
\begin{eqnarray}
\frac{d^4\Omega}{d\sigma^4}\left(\sigma,\Delta(\sigma)\right) & = & 0,
\label{trimix}
\end{eqnarray}
which is the analog of the condition $2 B(T^2-\mu^2)-N_c=0$ for the pure
chiral case.  We have solved Eq.~(\ref{trimix}) numerically to determine
$(\mu_{3\rm m},T_{3 \rm m})$ in Figs.~\ref{f:panel5} and~\ref{f:panel6}.

\begin{figure}[ht]
\epsfxsize=\figsize
\centerline{
\epsffile{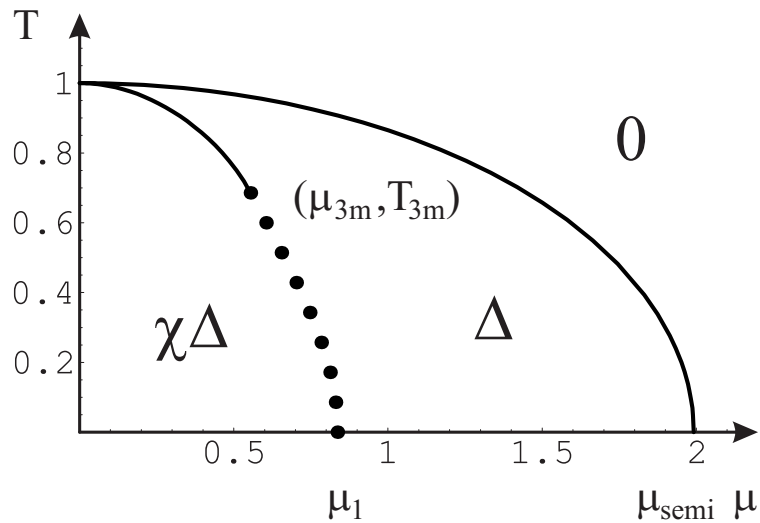}}
\vspace{.5cm}
\caption{Phase diagram for the maximal ratio $B/A=1.5$ that can be
  realized by the random matrix interactions. With respect to Fig.~5, the
  thermodynamic competition  favors the mixed broken symmetry phase
  at the expense of the chiral phase.}
\label{f:panel5b}
\end{figure}

{\bf Disappearance of the chiral phase.} ($\alpha_2(N_c) \le B/A$; see
Figs.~6 and~10).  The wedge of mixed broken symmetry gains space in the
$(\mu,T)$ plane at the expense of the chiral phase as $B/A$ increases above
$\alpha_2(N_c)$. The chiral phase eventually shrinks to a vertical line along
the $\mu=0$ axis when $B/A=N_c/2$, as shown in Fig.~\ref{f:panel5b}. That the
chiral phase is still present for $\mu=0$ can be understood as follows. When
$B/A =N_c/2$ and $\mu = 0$, both condensation fields appear in
Eq.~(\ref{Omega}) in the combination $\sigma^2+\Delta^2$. A pure diquark
solution $(\sigma,\Delta)=(0,\Delta_0)$ can thus always be rotated into a
pure chiral solution $(\sigma,\Delta)=(\Delta_0,0)$~\cite{VanJac99}, and thus
does not represent an independent phase. This symmetry is however no longer
present as soon as $\mu \neq 0$, in which case the diquark phase becomes
thermodynamically independent from the chiral phase.

Recall that the ratio $B/A = N_c/2$ is the maximum ratio that our model can
realize~\cite{VanJac99}. It is instructive however to explore higher coupling
ratios, although they do not necessarily describe physical situations, by
setting $B/A > N_c/2$ by hand in Eq.~(\ref{Omega}). We find that higher $B/A$
force the diquark phase to grow in size at the expense of the mixed broken 
symmetry phase. Figure~\ref{f:panel6} shows one example with $B/A =1.8$ and
$N_c = 3$. The similarities between the critical properties of the mixed
broken symmetry phase and those of the pure chiral broken phase are now
clear: Compare for instance the critical lines between Figs.~\ref{f:panel1}
and~\ref{f:panel6}.

\section{Discussion}
\label{s:discuss}

{ \bf Baryon density discontinuity.} We have argued that the potential
$\Omega$ in Eq.~(8) represents the non-analytic contribution to the
thermodynamics which is directly associated with the breaking of chiral and
color symmetry. The potential $\Omega$ should therefore not be used too
literally in computations of the bulk properties of the phases we have
encountered. It may however give reasonable estimates of discontinuities near
a phase transition. For instance in the pure chiral case, the random matrix
models of Ref.~\cite{HalJac98} estimate that the baryon density $n_B = -(1/3)
\partial \Omega(\mu,T)/\partial \mu$ changes discontinuously at the first
order point along $T=0$ by an amount $\Delta n_B \sim 2.5\, n_0$. Here, $n_0
= 0.17\, {\rm fm}^{-3}$ is the density of normal nuclear matter. The result
for $\Delta n_B$ relies on an evaluation of the number of degrees of freedom
$N$ from instanton models and seems to be a reasonable estimate of the
baryon density discontinuity~\cite{HalJac98}.

$\Delta n_B$ is modified by the presence of diquark condensation.  We
consider single-gluon exchange with $N_c=3$, which realizes $B/A=3/4$, and
work in the limit $T=0$. Taking the derivative of Eq.~(\ref{DeltaOmega}) with
respect to $\mu$, we find the discontinuity in baryon density at the point
$\mu_1$ of equal pressure between chiral and color phases to be
\begin{eqnarray}
N_\Delta-N_\chi & = & \frac{2}{\mu_1} \left(1+\frac{7}{3} B \mu_1^2\right).
\label{deltaNcol}
\end{eqnarray}
Here, $\mu_1$ obeys $\Omega_\chi=\Omega_\Delta$ in Eq.~(\ref{DeltaOmega}); we
have $\mu_1 = 0.87 /\sqrt{B}$. Hence, $N_\Delta - N_\chi \sim 6.4
\,\sqrt{B}$ in the appropriate unit of inverse chemical potential.  Were no
diquark condensation to occur, we would have found a discontinuity
\begin{eqnarray}
N_0-N_\chi & = & \frac{6}{\mu_1} \left( 1+ \frac{B \mu_1^2}{3} \right),
\end{eqnarray}
where $\mu_1$ is now the point of equal pressure between chiral and trivial
phases. We want to keep the same $B$ as in Eq.~(\ref{deltaNcol}) so as to
compare two situations which realize the same vacuum chiral field $\sigma_0 =
\sqrt{3/B}$. The condition of equal pressure in Eq.~(\ref{mu1}) gives then
$\mu_1=0.914/\sqrt{B}$ and we have $N_0-N_\chi \sim 8.4 \,\sqrt{B}$. Thus, we
find that diquark condensation reduces the discontinuity in baryon density by
roughly twenty five percent.

{\bf Away from the chiral limit.} We now turn to study the effects of a small
quark current mass $m$ in Eq.~(\ref{Omega}). For $m \neq 0$, chiral symmetry
is explicitely broken, and the chiral condensate $\langle \psi^\dagger
\psi\rangle$ ceases to be a good order parameter. This affects the phase
diagrams in Figs.  1-6 in a number of ways.  We consider the effect of a
small mass $m$ chosen so that $m \sim 10$ MeV in units for which the vacuum
chiral field is $\sigma \approx \sqrt{3/B} \sim 400$ MeV and illustrate a
few cases in Figs.  11-14.

Figure 11 shows the limit of small ratios $B/A$ which favor chiral over
diquark condensation. Since $\langle \psi^\dagger \psi \rangle $ is no longer
a good order parameter, any two given points in the phase diagram can be
connected by a trajectory along which no thermodynamic discontinuity occurs.
It results that the second-order line $L_{\chi,2}$ in
Eq.~(\ref{sigma_second}) is no longer present when $m \neq 0$. There remains,
however, a first-order line, which ends at a regular critical point
$(\mu_c,T_c)$. This point can be  located as follows.  Along the first-order
line, the potential $\Omega(\sigma,0)$ has two minima of equal depth
separated by a single maximum.  All three extrema become degenerate at the
critical point $(\mu_c,T_c)$, past which $\Omega(\sigma,0)$ possesses only
one minimum.  The location of $(\mu_c,T_c)$ can thus be determined from the
condition that $\Omega(\sigma,0)$ scales as
\begin{eqnarray}
\Omega(\sigma,0) - \Omega(0,0) \sim (\sigma - \sigma_0)^4,
\end{eqnarray}
at $(\mu_c,T_c)$ and for small deviations $|\sigma-\sigma_0|$. The location
of $(\mu_c, T_c)$, as well as $\sigma_0$, are then determined by requiring
the first three derivatives of $\Omega(\sigma,0)$ to vanish at the critical
point and for $\sigma = \sigma_0$.  A small mass tends to increase the
pressure in the low density `phase' with respect to that in the high density
`phase'. Its results that a mass $m$ displaces the first-order line
$L_{\chi,1}$ of Eq.~(\ref{Lchi1}) to higher $\mu$ by an amount linear in $m$.
The pressure increase also delays the onset of diquark condensation; the
ratio $B/A$ for which the diquark appears first increases linearly with $m$.

Figure 12 shows the case of QCD: there is now a second-order line
$L_{\Delta,2}$ which separates the high temperature phase from a mixed broken
symmetry phase; the dominant effect of $m$ on the diquark phase in Fig.~2 is
to produce a small chiral field $\sigma \sim m$. The effect on the
thermodynamics in both the chiral and the trivial phases is, however,
second order in $m$, and so is the displacement of the second-order line
$L_{\Delta,2}$ from Fig.~2 to Fig.~12.

The evolution of the phase diagram for higher $B/A$ parallels the evolution
we have outlined for $m=0$ in Figs. 3-6. The phase with a finite diquark
field grows until it eventually reaches the critical point $(\mu_c,T_c)$.
Higher ratios lead to a phase structure in which the first- and second-order
lines intersect as shown in Fig. 13. A wedge of another mixed broken symmetry
phase, initially with a large chiral field and a small diquark field, appears
at still higher $B/A$ on the left of the first-order line. This wedge grows
in size until it encounters the boundary of the other mixed broken symmetry
phase.  The second-order lines then merge into a single continuous line as
shown in Fig.~14.  This line is the locus of point for which
\begin{eqnarray}
\left. \frac{\partial^2 \Omega(\sigma,\Delta)}{\partial
    \Delta^2}\right|_{\Delta = 0} = 0,
\end{eqnarray}
and on which a chiral solution with a vanishing diquark field turns into a
saddle-point of $\Omega(\sigma,\Delta)$. Inside this boundary, the global
minimum of $\Omega(\sigma, \Delta)$ describes a single mixed broken symmetry
phase. This phase again exhibits properties similar to those of a chiral
phase. It contains in particular a first-order line which ends at a critical
point $(\mu_{\rm cm},T_{\rm cm})$ at which the potential $\Omega$ scales as
$\Omega(\sigma, \Delta(\sigma)) - \Omega(0,0) \sim (\sigma-\sigma_0)^4$,
where $\Delta(\sigma)$ is the diquark field for fixed $\sigma$.

To summarize, the effects of a small mass is linear for the chiral and mixed
broken symmetry phases and quadratic for the diquark and trivial phases. The
chiral field no longer represents a good order parameter, and the second-order
lines of vanishing second derivatives with respect to $\sigma$, i.e.
$L_{\chi,2}$ in Eq.~(\ref{sigma_second}) and $L_{\rm mix}$ in
Eq.~(\ref{mixline}), disappear. The tricritical points become regular critical
points while the tetracritical point in Fig.~5 disappears.

{\bf The $N_c=2$ and $N_c \to \infty$ limits.} In order to make connection
with known results and to obtain some insight on the dependence of the phase
structure on the number of colors, it is interesting to consider the limits
$N_c = 2$ and $N_c \to \infty$.  First of all when $N_c=2$, the Hermitian
matrix models realize only a single coupling ratio $B/A = 1$~\cite{VanJac99}. We
focus on single-gluon exchange for which the potential in
Eq.~(\ref{Omega}) becomes
\begin{eqnarray}
\Omega_{2}(\sigma,\Delta) & = & A (\sigma^2 +\Delta^2) -
\log[(\sigma-\mu)^2+\Delta^2+T^2] - \log[(\sigma+\mu)^2+\Delta^2+T^2].
\label{Om2}
\end{eqnarray}
A rotational symmetry appears at $\mu =0 $ as $\Omega_2$ depends on chiral
and diquark fields via the combination $\sigma^2+\Delta^2$.  Diquark and
chiral phases do not in this case represent independent states. We find for
the combined fields that 
\begin{eqnarray}
\sigma^2 + \Delta^2 & = & {2 \over A} - T^2,
\end{eqnarray}
below a critical temperature $T_c = \sqrt{2/A}$, while symmetry is restored
above $T_c$. This rotational symmetry is a consequence of the pseudo-reality
of $SU(2)$-QCD, a property by which the Dirac operator
$D = i \sum_{\mu a} \gamma^\mu \lambda^a A_{\mu a} + m$ commutes with $\tau_2
C\gamma^5 K$ where $\tau_2$ is the antisymmetric $2\times 2$ color matrix and
$K$ the complex conjugation operator. For $\mu = 0$, this property permits
one to arrange color and flavor symmetries into a higher $SU(4)$ symmetry,
see for instance~\cite{DiaPet92,SmiVer95,KogSte99}.

At finite $\mu$, however, the $SU(4)$ symmetry is explicitly broken. 
The global minimum of $\Omega(\sigma,\Delta)$ always has $\sigma = 0$, and
the system prefers diquark condensation over chiral symmetry breaking.
This results agrees with instanton models~\cite{CarDia99,DiaFor96}, and
lattice calculations~\cite{DagKar86}.  We now have a second-order phase
transition from an ordered state with $\Delta^2 = 2/A -\mu^2 - T^2$ in the
low $T$ and $\mu$ region to a symmetric phase at high $T$ and $\mu$.

The opposite limit $N_c \to \infty$ is more subtle. In microscopic models, it
is expected on general grounds that the quark-quark interaction is suppressed
with respect to the $\bar qq$ channel by powers of
$1/N_c$~\cite{DiaFor96,tHo73}.  In the present model, we observe that diquark
condensation disappears as $N_c \to \infty$ if the interaction is
single-gluon exchange. Its coupling ratio $B/A = N_c/(2 N_c - 2) \to 1/2$ as
$N_c \to \infty$. Thus, we have $B/A \ll 0.139 N_c$ and the only possible
topology for the phase diagram is that in Fig.~1. Hence, no diquark
condensate forms. Other Hermitean interactions with $B/A \sim O(N_c)$ can
however explore the full range of phase diagrams which display diquark
condensation. The actual number of possible toplogies is however reduced to
five, as $\alpha_2(N_c)\to N_c/\sqrt{8}$ for $N_c \to \infty$ (see Appendix)
and Fig.~4 can no longer be realized.

We summarize the variation with $N_c$ of the coupling ratios characterizing a
change in topology and of the ratio realized by single-gluon exchange in
Table 1.

{\bf Comparison with a microscopic model:} In contrast to microscopic models,
the random matrix interaction does not lead to a logarithmic instability of
the gap equation near the Fermi surface. To clarify this effect, we consider
diquark condensation in a chiral symmetric phase at $T=0$ and
compare the random matrix approach to a microscopic model. We  choose
here the NJL study of Berges and Rajagopal~\cite{BerRaj99}.  In the random
matrix formulation  the gap equation
\begin{eqnarray}
2 A \Delta & = & {\Delta \over \Delta^2 + \mu^2}
\label{BCSgapRMM}
\end{eqnarray}
has a non-trivial root $\Delta \sim \sqrt{\mu_c - \mu}$ with
$\mu_c = \sqrt{2/A}$. The model of~\cite{BerRaj99} gives an equation of the
form
\begin{eqnarray}
2 A' \Delta & = & \Delta \int_0^\infty dq q^2 F^4(q) \left({1\over
    \sqrt{(q-\mu)^2+\Delta^2 F^4(q)}}+
{1\over \sqrt{(q+\mu)^2+\Delta^2 F^4(q)}}\right),
\label{BCSgap}
\end{eqnarray}
where $F(q)$ is an appropriate form factor which falls off over $q \sim
\Lambda \sim O(\Lambda_{\rm QCD})$. Clearly, the right side of
Eq.~(\ref{BCSgap}) contains a singularity for $q \sim \mu$ and $\Delta \to 0$
which is absent in Eq.~(\ref{BCSgapRMM}). This singularity has two
consequences. First, the behavior of the diquark field for large $\mu$
depends sensitively on the form factor. As $\Delta \to 0$, the singularity at
$q \sim \mu$ in the right side of Eq.~(\ref{BCSgap}) gives $ 2 A' \approx
\mu^2 F^4(\mu) \log \Lambda \mu /\Delta^2$ to logarithmic order. Thus,
instead of the square root behavior $\Delta(\mu)\sim (\mu_c - \mu)^{1/2}$ of
the random matrix approach, $\Delta$ now vanishes as $\mu \to \infty$ as
$\Delta(\mu) \propto \exp(- c/(\mu^2 F^4(\mu)))$, where $c$ is a
constant.~\footnote{This result is only true for a smooth cutoff $F(q)$.  For
  a sharp cutoff $F(q)=\Theta(\Lambda - \mu)$, where $\Theta(x)$ is the
  Heavyside function, the diquark field exhibits the square root
  singularity.} This tail is exponentially sensitive to form factors; with
regards to establishing the general topology of the phase diagram, such
behavior is not qualitatively different from $\Delta = 0$.  The second
consequence of the logarithmic singularity in Eq.~(\ref{BCSgap}) is that
$\Delta(\mu)$ must be non-monotonic for intermediate $\mu$.  This profile can
be seen by drawing in the plane $(\Delta,\mu)$ the lines of constant height
for the right side of Eq.~(\ref{BCSgap}): these lines must go around the
point $(\mu_0,0)$ where the logarithmic singularity is the strongest as
$\Delta \to 0$, and $\Delta(\mu)$ reaches a maximum at $\mu_0$.

{\bf Color dependency in the chiral fields.} The patterns of symmetry
breaking which we have described can become even richer if we allow the
chiral fields to depend on color. This possibility arises in the instanton
model of Carter and Diakonov~\cite{CarDia99}, who remarked that the
Dyson-Gorkov equations close in color space on the condition that gapped and
ungapped quarks can develop different masses. We studied the effects of this
additional degree of freedom in the limit of zero chemical
potential~\cite{VanJac99} and found no change in the phase structure for
ratios $B/A \le N_c/2$. The situation is different for finite $\mu$.
Choosing different masses for the two gapped and the $N_c-2$ ungapped quarks
leads in some limits to an increase in the pressure of phases with finite
diquark fields. The main consequence is an increase of both the parameter
range and the region in the $(\mu,T)$ plane for which the mixed broken
symmetry phase exists. This result is obvious since the mixed broken symmetry
phase is the only phase which can exploit this additional degree of freedom.
However, there is also a general mechanism at work by which a change of phase
structure always takes place above a certain threshold for the coupling
constants.

We wish to illustrate these effects in a few cases and concentrate on the
limit $N_c =3$ and $m = 0$ for clarity. We denote the chiral fields for color
$1$ and $2$ by $\sigma_1$ and that for color $3$ by $\sigma_3$. In order to
permit these fields to be different, we include the projection of the
interaction onto a chiral-$\lambda_8$ channel as described
in~\cite{VanJac99}. For $\mu\neq 0$, the thermodynamical potential becomes
\begin{eqnarray}
\Omega (\sigma_1,\sigma_3,\Delta) &=& A \Delta^2 + B(\beta_1
\sigma_1^2+\beta_2 \sigma_1 \sigma_3 + \beta_3 \sigma_3^2)  
- \log [(\sigma_1 - \mu)^2 + \Delta^2 + T^2]      
- \log [(\sigma_1 + \mu)^2 + \Delta^2 +  T^2] \nonumber \\
&& - {1 \over 2} \log [ (\sigma_3 -\mu)^2 + T^2 ] -{1\over 2} \log[(\sigma_3 +
\mu)^2 + T^2],
\label{Omegasplit}
\end{eqnarray}
where 
\begin{eqnarray} 
\beta_1 = {4\over 9} + {C \over 3 B}, \quad
\beta_2 = {4 \over 9} - {2 C \over 3 B}, \quad
\beta_3 = {1 \over 9} + {C \over 3 B},
\label{beta123}
\end{eqnarray}
describe the coupling between $\sigma_1$ and $\sigma_3$. $C$ is a coupling
constant associated with the chiral-$\lambda_8$ channel~\cite{VanJac99}. By
fine tuning the various Lorentz and color channels which compose the random
matrix interactions, we can realize a range of ratios $B/C$ for any fixed
$B/A$.~\footnote{This range is $-3/16 \le B/C \le 3/2$ for $B/A < 3/4$ and
  then reduces linearly to $B/C = 3/2$ for $B/A=3/2$.} New patterns of
symmetry breaking only appears when $C > 0$ and the chiral-$\lambda_8$
channel is attractive~\cite{VanJac99}.

Consider first single-gluon exchange. The coupling ratios are $B/A = 3/4$ and
$C/B = -3/16$. The chiral-$\lambda_8$ channel is repulsive, and the phases
with the largest pressure always satisfy $\sigma_1 = \sigma_3$, which is the
case shown in Fig.~2. If we now fine tune the interactions so as to increase
$B/C$ to $B/C \sim 3/2$ while keeping $B/A = 3/4$, the phase diagram changes
substantially. The first-order line between the chiral and trivail phases
splits at the point where it meets the diquark transition line into two
first-order lines. Together with the $T=0$ axis, they delimitate a wedge of
mixed broken symmetry phase with $\sigma_1 \neq \sigma_3$ and $\Delta \neq
0$.  Thus, with an attractive channel $B/C \sim 3/2$ the mixed broken
symmetry phase appears much earlier than previously discussed.  This is an
extreme case. For fixed $B/A$, the mixed broken symmetry phase does not
appear immediately as $B/C$ increases from $0$: there is a threshold value
above which the new phase appears (this value is $B/C \sim 1.29$ for
$B/A=0.75$).  Therefore, it takes large variations of the coupling constants
$B/A$ and $B/C$ away from the values expected for single-gluon exchange to
modify the phase structure of Fig.~2.

To complete the picture of the effects of an attractive channel $C>0$,
consider next $B/A = 3/2$. This ratio can only be realized by a color
diagonal interaction for which $B/C = 3/2$~\cite{VanJac99}. In this case,
there is no freedom in fine tuning $B/C$ to other values. The effect of
splitting masses in colors is now maximal. $\beta_2$ in Eq.~(\ref{beta123})
vanishes and $\sigma_3$ decouples from the other two fields. Its gap equation
leads to the same solutions as a pure chiral phase with $\sigma \neq 0$ and
$\Delta =0$ and the phase diagram for the third color is that of Fig.~1. For
$B/A = B/C =3/2$, the partial pressure for colors $1$ and $2$ has the same
form as in the limit $N_c = 2$, see Eq.~(\ref{Om2}). There is then a
rotational symmetry between $\sigma_1$ and $\Delta$ for $\mu = 0$ and
$\sigma_1$ vanishes for $\mu \neq 0$ while $\Delta = 2/A -\mu^2 -T^2$ for
$\mu^2+T^2 \le 2/A$.  The overall phase diagram is obtained by superposing the
two pictures; we have a mixed broken symmetry phase with $\sigma_1=0$,
$\sigma_3 \neq 0$, and $\Delta \neq 0$ on the left of the first-order line
$L_{\chi,1}$ of Eq.~(\ref{Lchi1}). This phase is contiguous to a pure diquark
phase which develops as in Fig.~2 on the right of $L_{\chi,1}$ and below the
semicircle. With respect to Fig.~5, the mixed broken symmetry phase thus
occupies a wider area in the $(\mu,T)$ plane.

{\bf A color-$6$ condensate.} It has been suggested that an energy gain may
result if the third color condenses in a spin-$1$ color symmetric
state~\cite{AlfRaj98}. We find no such condensation for single-gluon
exchange.  We can however fine tune the interactions so as to keep $B/A =3/2$
and allow this channel to develop. The Fierz constant for projecting on a
color $6$ is half that for a $\bar 3$ color state. Thus, we can
understand qualitatively how a color-$6$ behaves by repeating our previous
analysis with an additional phase which now may develop inside a semicircle
of radius $\sqrt{2}$ smaller than the radius of the diquark phase.  This
semicircle crosses the first-order line beween chiral and diquark phases at
$B/A \sim 0.754$ and a color-$6$ phase can, in favorable cases, increase the
pressure of the system.  However, as far as QCD is concerned, a color-$6$
condensate should be very small since its threshold ratio is very close to
$B/A = 3/4$.  This is an example of a result which cannot be regarded as
robust: in a microscopic theory, the fate of the color-$6$ phase will
inevitably depend on the details of the interaction and on whether these give
rise to exponential tails for the associated condensation field. By contrast,
the chiral and diquark phases are fully developed for $B/A =3/4$ and the
phase diagram in Fig.~2 should thus be considered robust.

\section{Conclusions}
\label{s:conclusions}

Our random matrix model leads to a thermodynamic potential which has a very
simple form. Yet, it contains enough physics to illustrate in a clear way the
interplay between chiral symmetry breaking and the formation of quark Cooper
pairs.  We have found that this interplay results in a variety of phase
diagrams which can be characterized by a total of only six different
topologies. Single-gluon exchange leads to the topology shown in Fig.~2, a
phase diagram familiar from microscopic models.  We have considered the
chiral and scalar diquark channels, which seem the most promising ones. We
have found that chiral and diquark phases are fully developped for the ratio
realized by single-gluon exchange and that it takes large variations in the
coupling ratios $B/A$ and $C/B$ to depart from that result.  On the other
hand, other less attractive condensation channels seem sensitive to coupling
constant ratios and are expected in microscopic models to depend on the
details of the interactions.  Furthermore, these channels develop 
 weak condensates at best. It is worth keeping in mind that the present picture
is mean-field; we expect that quantum fluctuations will inevitably have large
effects on the weak channels, which should thus not be considered robust.  We
 conclude that QCD with two light flavors should realize the topology
suggested by single-gluon exchange and that this topology is stable against
variations in the detailed form of the microscopic interactions.

Many effects lie in the mismatch between the number of colors involved in
each order parameter: all $N_c$ colors contribute to the chiral field, while
a Cooper pair involves two colors. A result of this for $N_c = 3$ is that the
chiral broken phase prevails at low densities. Furthermore, our model
reproduces to a reasonable extent the expected limits at small and large
$N_c$.  We expect such counting arguments to be valid in both microscopic
models and lattice calculations.  More generally, we believe that random
matrix models can provide insight into calculations of QCD at finite density
by providing simple illustrations for many of the mechanisms which are at
work.

Not all the phase diagrams that we have studied are directly relevant to QCD.
However, many of their characteristics such as the presence of tricritical
and tetracritical points are generic to systems in which two forms of order
compete. Our model could naturally be extended to the study of nuclear or
condensed matter systems in which such competion takes place. The
construction of a genuine theory seems technically involved at first
glance~\cite{VanJac99} but in fact contains only three basic ingredients: the
identification of the symmetries at play and their associated order
parameters, the knowledge of the elementary excitations in a background of
condensed fields, and the calculation of the range of coupling constants
realized by the random matrix interactions. These three components are
sufficient to build a thermodynamical potential and determine the resulting
phase structures in parameter space.

\section*{Acknowledgements}

We are grateful for stimulating discussions with G. Carter, D. Diakonov, H.
Heiselberg, and K. Splittorff.

\appendix
\section{Calculation of $\alpha_1(N_c)$ and $\alpha_2(N_c)$}
\label{a:alpha1}

In this appendix, we calculate the ratios $\alpha_1(N_c)$ and
$\alpha_2(N_c)$, which respectively characterize the onset of the
$\chi\Delta$-phase and the appearance of the tetracritical point.

\subsection{The ratio $\alpha_1(N_c)$.}

The mixed broken symmetry phase appears first for the coupling ratio $B/A$
for which the instability line $L_{\rm mix}$ of Eq.~(\ref{mixline}) crosses
the first-order line between chiral and diquark phases. This crossing occurs
on the $T=0$ axis. The line $L_{\rm mix}$ meets the $T=0$ axis at $\mu =
\mu_{\rm mix}$, Eq.~(\ref{mumix}),
\begin{eqnarray}
\mu_{\rm mix}^2 = \frac{N_c^2 A}{4 B^2} - \frac{N_c}{2 B}, \nonumber
\end{eqnarray}
while the condition of equal pressure at that point gives
\begin{eqnarray}
\Omega_\Delta - \Omega_\chi = (N_c - 2)+(A+B) \mu_{\rm mix}^2-2 \log \frac{A
  \mu_{\rm mix}^2}{2}+ N_c \log \frac{B \mu_{\rm mix}^2}{N_c} = 0.
\end{eqnarray}
Combining these two equations gives the determining equation for
$\alpha_1 = B/A$,
\begin{eqnarray}
(N_c-2) & + & \frac{N_c}{2 \alpha_1^2} \left(\frac{N_c}{2}-\alpha_1\right) 
\left(1+\alpha_1\right) 
- 2 \log \left[\frac{N_c}{4 \alpha_1} \left(\frac{N_c}{2 \alpha_1}-1\right)
\right] 
+ N_c \log \left[\frac{1}{2}\left(\frac{N_c}{2 \alpha_1} -1\right)\right] = 0.
\end{eqnarray}
This is a transcendental relation which needs to be solved numerically for
each $N_c$. In particular, we find 
 $\alpha_1(2)=1$, $\alpha_1(3)=1.05$, 
and $\alpha_1(N_c \to \infty)\sim 0.321 N_c$.

\subsection{The ratio $\alpha_2(N_c)$.}

Two conditions determine $\alpha_2(N_c)$. Coming from small $B/A$, 
$\alpha_2(N_c)$ corresponds to the ratio at which the wedge of the
$\chi\Delta$-phase reaches the critical point $(\mu_4,T_4)$ of
Eq.~(\ref{tetracritical}). Decreasing $B/A$ from large values,
$\alpha_2(N_c)$ marks the coexistence of the tetracritical point
$(\mu_4,T_4)$ and the tricritical point $(\mu_{3\rm m},T_{3\rm m})$. We now
show that these two conditions are equivalent, and thus that the transition
from Fig.~4 to Fig.~5 is continuous. 

To proceed, we concentrate on the region
near $(\mu_4,T_4)$ where we perform a small
field expansion  of the potential $\Omega(\sigma,\Delta)$ in
Eq.~(\ref{Omega}),
\begin{eqnarray}
\Omega(\sigma,\Delta) & \approx & \Omega(0,0) + a_0\, \sigma^2 + b_0\, \Delta^2+
\frac{a_1^2}{2}\, \sigma^4 + \frac{b_1^2}{2}\, \Delta^4 +c_1 \,\sigma^2
\, \Delta^2 + {\cal O} ({\rm min}^6(\Delta,\sigma)).
\label{Omegaexpanded}
\end{eqnarray}
Here, the coefficients of the quadratic terms are linear in $\delta \mu = \mu
-\mu_4$ and $\delta T = T-T_4$ and vanish at $(\mu_4,T_4)$, while those of
the quartic terms are
\begin{eqnarray}
a_1^2  = \frac{2 B^2}{N_c}-\frac{A^2 N_c}{4},\quad
b_1^2 = \frac{A^2}{2}, \quad
c_1 =\frac{2 A B}{N_c} -\frac{A^2}{2},
\label{a1b1c1}
\end{eqnarray}
to leading order in $\delta\mu$ and $\delta T$.

The condition that the $\chi\Delta$-phase includes $(\mu_4,T_4)$ can be
determined as follows. While $B/A$ increases from below $\alpha_2(N_c)$, the
tip of the wedge of the $\chi\Delta$-phase slides along the the line of equal
pressures between chiral and diquark phases, which we denote by $L_{\chi \to
  \Delta,1}$.  Minimizing $\Omega(\sigma,\Delta)$ in
Eq.~(\ref{Omegaexpanded}) to find the respective pressure in the $\chi$- and
$\Delta$-phases, this line is determined near $(\mu_4,T_4)$ by
\begin{eqnarray}
L_{\chi\to \Delta,1} \approx \frac{b_0^2}{b_1^2} - \frac{a_0^2}{a_1^2} = 0.
\label{firstnear4}
\end{eqnarray}
The left boundary of the $\chi\Delta$-phase is the instability line 
\begin{eqnarray}
L_{\rm mix} \equiv 
\left.
\frac{\partial^2 \Omega}{\partial \Delta^2}
\right|_{\Delta = 0,\sigma = \sigma_0} = 0,
\end{eqnarray}
where $\sigma_0$ is the $\sigma$ field in the chiral phase. From Eq.~(\ref{Omegaexpanded}), this gives near $(\mu_4,T_4)$
\begin{eqnarray}
L_{\rm mix} \approx b_0 - \frac{a_0}{a_1^2}{c_1}=0.
\label{Lmixnear4}
\end{eqnarray}
The tip $P_{\rm mix}$ of the $\chi\Delta$-phase is the intercept of $L_{\rm
mix}$ and $L_{\chi\to\Delta,1}$. From Eqs.~(\ref{firstnear4}) and
(\ref{Lmixnear4}) $P_{\rm mix}$ thus obeys
\begin{eqnarray}
a_0^2 \left(c_1^2 - a_1^2 b_1^2 \right) = 0.
\label{alpha2eq}
\end{eqnarray}
Now, imagine that $B/A$ is strictly smaller but near
$\alpha_2(N_c)$.  Then, $P_{\rm mix} \neq (\mu_4,T_4)$, and we have $a_0 \neq
0$. The location of $P_{\rm mix}$ is therefore determined by the condition
$a_1^2 b_1^2 = c_1^2$, where the coefficients $a_1$,
$b_1$ and $c_1$ are those of Eq.~(\ref{a1b1c1}) augmented by linear
corrections in $\delta \mu$ and $\delta T$. If we now let $B/A \to
\alpha_2(N_c)$, the tip $P_{\rm mix}$ reaches $(\mu_4,T_4)$ and the linear
corrections in $a_1$, $b_1$, and $c_1$ vanish. Using Eq.~(\ref{a1b1c1}), and
solving $a_1^2 b_1^2 = c_1^2$ for $B/A$ gives
\begin{eqnarray}
\alpha_2(N_c)={B \over A} \equiv \frac{4 N_c - N_c^{3/2} 
\sqrt{2 N_c-4}}{4 (4 - N_c)}, \nonumber
\end{eqnarray}
which is the result stated in Eq.~(\ref{alpha2}).

Coming from large ratios $B/A$, the determining condition for $B/A =
\alpha_2(N_c)$ requires that $(\mu_4,T_4)$ obeys Eq.~(\ref{trimix}),
\begin{eqnarray}
\frac{d^4\Omega}{d\sigma^4}(\sigma,\Delta(\sigma)) & = & 0, \nonumber
\end{eqnarray} 
where $\Delta(\sigma)$ is the diquark field in the $\chi\Delta$-phase. Taking
the derivative of Eq.~(\ref{Omegaexpanded}) with respect to $\Delta$, we find
$\Delta^2(\sigma) \approx - (c_1 \sigma^2 +b_0)/b_1^2$ which when inserted
back in Eq.~(\ref{Omegaexpanded}) gives
\begin{eqnarray}
\frac{d^4\Omega}{d\sigma^4}(\sigma,\Delta(\sigma)) & \approx & a_1^2 -
\frac{c_1^2}{b_1^2},
\end{eqnarray}
in the neighborhood of $(\mu_4,T_4)$. Setting the right side to zero, we
recover the previous condition of Eq.~(\ref{alpha2eq}).

\newpage

\vspace{2cm}

\vspace{2cm}

\vspace{2cm}

\vspace{2cm}

\vspace{2cm}

\begin{figure}[ht]
\epsfxsize=\figsize
\centerline{
\epsffile{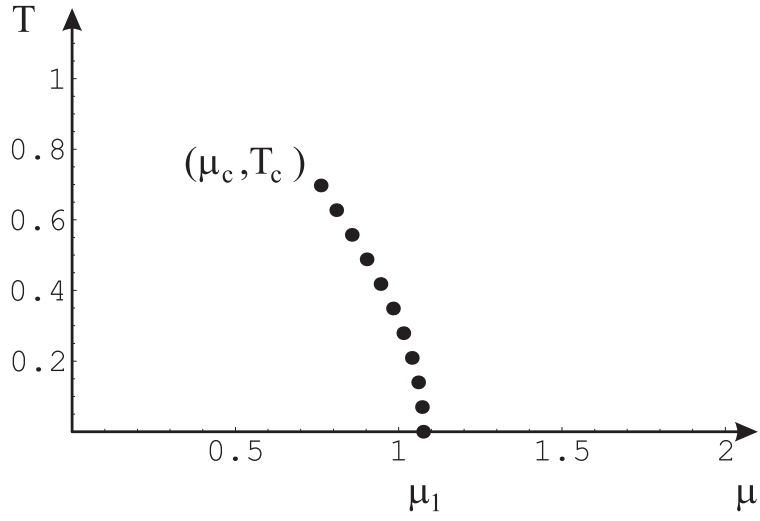}}
\vspace{.5cm}
\caption{Phase diagram for a ratio $B/A =0.2$ and a small quark mass
$m$.}
\label{f:fig1m}
\end{figure}

\vspace{2cm}

\begin{figure}[ht]
\epsfxsize=\figsize
\centerline{
\epsffile{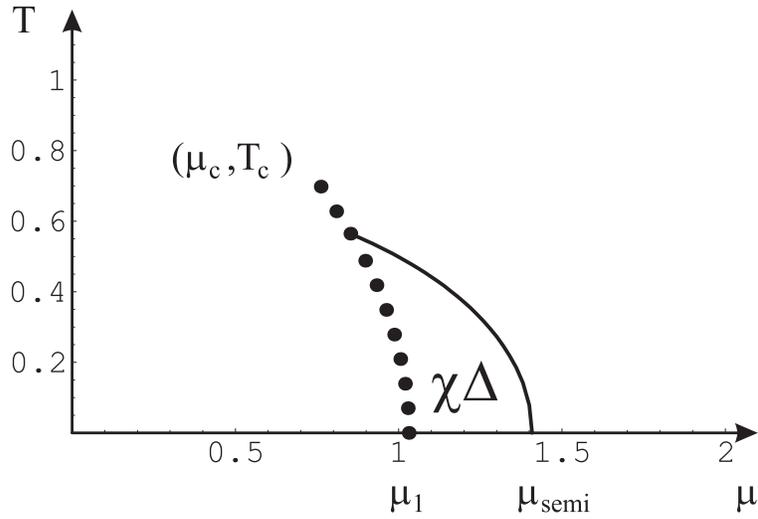}}
\vspace{.5cm}
\caption{ Phase diagram for the coupling ratio of single-gluon
exchange, $B/A = 0.75$, and a small quark mass $m \sim 10$ {MeV}.
The first-order line ends at a critical point $(\mu_c,T_c)$.}
\label{f:fig2m}
\end{figure}

\begin{figure}[ht]
\epsfxsize=\figsize
\centerline{
\epsffile{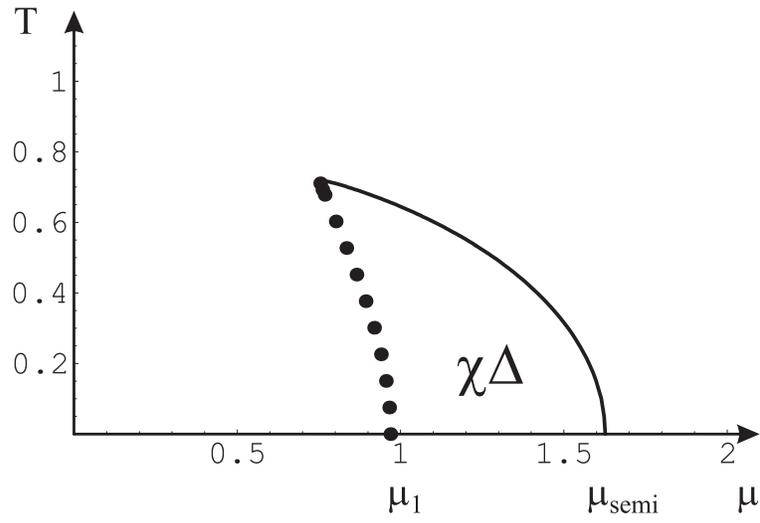}}
\vspace{.5cm}
\caption{Phase diagram for $B/A=1.0$ and a small quark mass $m \sim 10$
MeV.}
\label{f:fig3m}
\end{figure}

\vspace{2cm}

\begin{figure}[ht]
\epsfxsize=\figsize
\centerline{
\epsffile{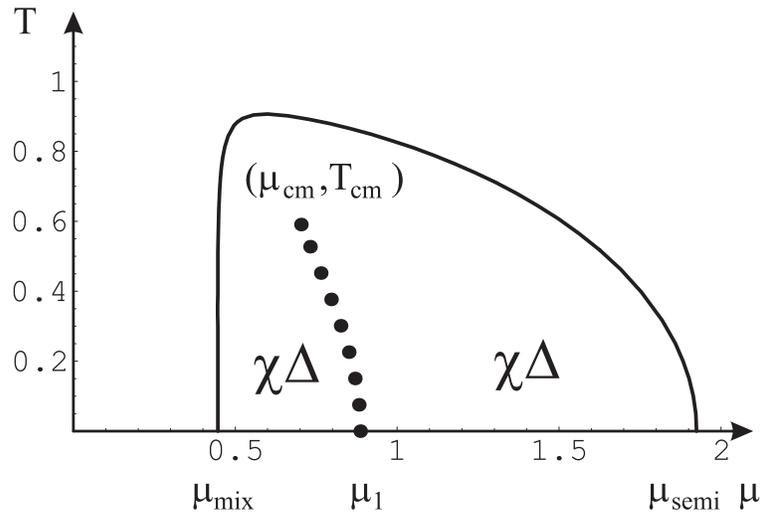}}
\vspace{.5cm}
\caption{Phase diagram for $B/A = 1.4$ and a small quark mass $m$. The
first-order line ends at a critical point $(\mu_{\rm cm},T_{\rm cm})$.}
\label{f:fig4m}
\end{figure}

\newpage

\begin{tabular}{l p{1em} r @{.} l p{1em} r @{.} l p{1em} r @{.} l l }
\multicolumn{11}{c}{Table 1. A few characteristic coupling ratios.} \\
\hline 
\hline
Ratio $B/A$ & &
\multicolumn{2}{c}{$N_c = 2$} & 
&
\multicolumn{2}{c}{$N_c = 3$} & 
&
\multicolumn{3}{l}{$N_c \to \infty$} \\
\hline
Onset of diquark condensation &  & \multicolumn{2}{l}{} & 
&
0 & 418 & 
& 
0 & 139 &  $N_c$ \\
$\alpha_1(N_c)$ & & 1 & 0 &
& 
1 & 050 & 
&
0 & 321 & $N_c$ \\
Disappearance of the tricritical point. & & \multicolumn{2}{l}{} & 
&
1 & 061 &
&
0 & 354 & $N_c$ \\
$\alpha_2(N_c)$ & & 1 & 0 & 
&
1 & 163 & 
&
0 & 354 & $N_c$ \\
Maximum ratio $N_c/2$ &  & 1 & 0 & 
&
1 & 5 &
&
0 & 5 & $N_c$ \\
Single-gluon exchange & & 1 & 0 & 
&
0 & 75 & 
&
0 & 5  \\
\hline
\hline
\end{tabular}

\widetext
\end{document}